# Exploring High Thermal Conductivity Polymers via Interpretable Machine Learning with Physical Descriptors

**Xiang Huang[1,†], Shengluo Ma[1,†], C. Y. Zhao[1], Hong Wang[2], and Shenghong Ju[1, 2, *]**

[1] China-UK Low Carbon College, Shanghai Jiao Tong University, Shanghai, China

[2] Materials Genome Initiative Center, School of Material Science and Engineering, Shanghai Jiao Tong University, Shanghai, China

†Equal contribution

## ABSTRACT

The efficient and economical exploitation of polymers with high thermal conductivity is essential to solve the issue of heat dissipation in organic devices. Currently, the experimental preparation of functional thermal conductivity polymers remains a trial and error process due to the multi-degrees of freedom during the synthesis and characterization process. Polymer informatics, which efficiently combines data science, machine learning, and polymer experiment/simulation, leading to the efficient design of polymer materials with desired properties. However, available polymer thermal conductivity databases are rare, and establishing appropriate polymer representation is still challenging. In this work, we have proposed a high-throughput screening framework for polymer chains with high thermal conductivity via interpretable machine learning and physical-feature engineering. The polymer thermal conductivity datasets for training were first collected by molecular dynamics simulation. Inspired by the drug-like small molecule representation and molecular force field, 320 polymer monomer descriptors were calculated and the 20 optimized descriptors with physical meaning were extracted by hierarchical down-selection. All the machine learning models achieve a prediction accuracy $R^2$ greater than 0.80, which is superior to that of represented by traditional graph descriptors. Further, the cross-sectional area and dihedral stiffness descriptors were identified for positive/negative contribution to thermal conductivity, and 107 promising polymer structures with thermal conductivity greater than 20.00 W/mK were obtained. Mathematical formulas for predicting the polymer thermal conductivity were also constructed by using symbolic regression. The high thermal conductivity polymer structures are mostly π-conjugated, whose overlapping p-orbitals enable easily to maintain strong chain stiffness and large group velocities. The proposed data-driven framework should facilitate the theoretical and experimental design of polymers with desirable properties.

*Corresponding author: shenghong.ju@sjtu.edu.cn.

# 1. INTRODUCTION

Polymers are extensively used in industry and daily life, owing to various advantages of chemical inertness, mechanical flexibility and light weight [1]. As the organic electronics are becoming smaller while the power density keeps increasing, the thermal management and heat dissipation capability have attracted significant attention [2]. However, conventional polymers are thermal insulators with reported thermal conductivity in the range from 0.1 to 0.5 W/mK, preventing the development of organic electronics [3]. Polymers with high thermal conductivity are urgently demanded in organic energy storage and electronic devices to accommodate revolutionary innovations in organic electronics and optoelectronics [4]. The polymer morphology and topology were found to be closely related to thermal conductivity [5]. Increasing the crystallite orientation and crystallinity can significantly reduce the phonon scattering and enhance the thermal conductivity along the chain directions, which has been demonstrated by both experiments [6-8] and theoretical simulations [9-12]. A recent study has fabricated polyethylene (PE) films by disentanglement and alignment of amorphous chains with a metal-like thermal conductivity of 62 W/mK, over two orders of magnitude greater than that of classical amorphous polymers [6]. Moreover, molecular dynamics simulations have suggested that individual crystalline PE chains have a very high or even divergent thermal conductivity [11]. These findings provide opportunities for solving the heat dissipation problem of polymer devices.

Despite the fact that chain alignment, crystallinity, polymer fibers or even single-chain polymer structures exhibit great influence on the thermal characteristic [13-15], the polymer library is quite large, with as many as $10^8$ monomeric organic molecules known to exist in chemical space [16]. Current research on the thermal conductivity of polymers is still an Edisonian process, guided by intuition or experience in a trial-and-error approach that is time-consuming and expensive [17]. Moreover, most of the studies are conducted on simple structures such as PE [4,6,11], which makes it difficult to grasp the general rule of the factors affecting the thermal conductivity of polymers and to discover polymer molecular structures with high thermal conductivity in huge chemical space.

The field of polymer informatics [18], associated with the development of artificial intelligence and machine learning (ML) methods, attempts to utilize the data-driven centric method for physical property regulation or device development of organic materials to resolve the conflict between structural freedom and efficiency/cost in the traditional trial-and-error approach. The research on polymer informatics has attracted extensive attention and succeeded in recent years [19-21], involving the prediction of organic optical [22-24], electrical [25-27] and thermal properties [28-32]. Particularly, several efforts emerged in the search or design of structures with high TC as related to crystalline polymers [30], amorphous polymers [31,32] and copolymers [33]. Most of these studies have employed graph descriptors [31] or polymer chemistry fragment statistics [30,32,33] to describe monomer structures in informatics algorithms, also called fingerprints or representations. The graph descriptors generated rely on molecular/monomer graph information, formulated by knowledge domain feature engineering [34] or by attempting to form general descriptors [35]. Moreover, descriptors such as molecular access systems (MACCS) [36] are obtained through statistics of different chemical fragments, and are closely related to molecular graphs. Subsequently, they are collectively referred to as graph descriptors. The fingerprint is required for the unique, complete, minimal representation of each candidate, and the successful fingerprint is a challenging task [37]. Besides, polymers are composed of many repeating units, which are more complex than organic small molecules and require accurate capture of information on monomer connection sites [26]. The graph descriptions have long been applied and validated in the development of drug-like small molecules [38], and the availability of open-source toolkits such as RDKit [39] and Mol2vec [34] has facilitated their accessibility, which is also one reason that graph descriptions are popular in polymer informatics. However, the graph descriptor is in the form of a string of numeric vectors. The completeness of the molecular structure determines the coupling association between the digits. Hence, the relationship between molecular monomers and material properties is difficult to grasp.

Exploring the ensemble of physically independent descriptors for the representation of molecular structures is important in qualitative structure-property relationship (QSPR) modeling and enables more intuitive guidelines for molecular structure evaluation [40]. Feature engineering for the collection and reduction of physical descriptors are critical steps in determining effective capabilities in polymer informatics. The development of automatic, universal and efficient tools for the calculation of descriptors of organic molecules is of interest to researchers, which translates the chemical information encoded in the symbolic representation of molecules into useful numbers or some standardized experimental results [41]. Several open-source and commercial software [41-43] are available to calculate various types of molecular descriptors such as carbon atomic number, molecular weight and Extended Topochemical Atom (ETA) [44], which have been successfully applied in organic chemistry synthesis [45], molecular antibacterial activity prediction [46] and so on. In addition, the parameter conditions in experiments or simulations affect the molecular properties. For instance, the force-field-inspired descriptors such as types of bond, angle and dihedral have been validated for the prediction of the specific heat of polymers, even if the datasets are from experiments [29]. The number reduction of polymer features is another concern, as some descriptors may have little relevance to the target property, and a low-dimensional descriptor space is much easier to build up for the ML model [47]. Feature extraction and selection are the dominant approaches to reduce the dimensionality of features. Feature extraction creates subsets from the original data space, such as principal component analysis (PCA), where the specific meaning of the new features obtained is difficult to understand [48]. Feature selection retains the physical meaning of individual descriptors, while filters based on correlation evaluation have dependencies on mathematical models, like the Pearson and Spearman coefficients that consider the linear and monotonic relationships of the data, respectively [49]. Further, the filter methods do not involve ML models, which may lead to the inapplicability of the gained features. The wrapper-based feature selection techniques combine ML models to eliminate redundant features, including recursive feature elimination (RFE), sequential feature selection (SFS) and exhaustive feature selection (EFS) [50]. Testing different subsets of descriptors for informatics algorithms is the crucial feature of the wrapper approaches, and the key is the strategy of combining different descriptors. Typical RFE seeks to improve model performance by continuously reducing the low impact features from the remaining features in iteratively constructed ML models, which refer to the ranking of feature weights assigned by models such as random forests [48]. Thus, the RFE relies on the feature weight evaluation mechanism of the ML models.

Herein, focusing on the challenges of polymer monomer representation and feature selection, we proposed an ML interpretable framework integrated with high-throughput MD simulations for the discovery of polymer structures with high TC, as illustrated in Fig. 1. It consists of four components: 1) polymer library construction, 2) MD simulation for the TC of polymers; 3) monomer feature representation and hierarchical down-selection; 4) ML models construction for TC prediction. The training data was collected from the literatures [51,52], and candidates from the databases of PolyInfo [53] and PI1M [54] were applied for the virtual screening of high TC structures. All polymer monomers were identified by the SMILES (simplified molecular input line entry system) strings and formed one-dimensional polymer chains by replication. The TC of training datasets was calculated by MD simulations with the second generation of the general AMBER force field – GAFF2 [55]. Inspired by drug-like molecular representation and molecular force fields, we obtained 320 physical descriptors by mordred software [41] calculation and force field parameter file extraction, and retained 20 optimized descriptors by hierarchical down-selection. We then trained random forest (RF), extreme gradient boosting (XGBoost) tree-based models, and multilayer perceptron (MLP) neural network model separately to establish the relationship between the optimized descriptors and the TC of these benchmark polymer datasets. Further, we analyzed the feature importance of each optimized descriptor and extracted the chemical heuristic for high TC polymers design through SHAP analysis [56]. Using the trained ML models, 107 promising polymers with TC greater than 20.00 W/mK were identified, which are served for symbolic regression to derive mathematical formulas for expressing the TC of promising polymers. Last, we discussed the TC mechanisms of eight typical polymers. Overall, the proposed approach is beneficial for theoretical or experimental investigations of high TC polymers.

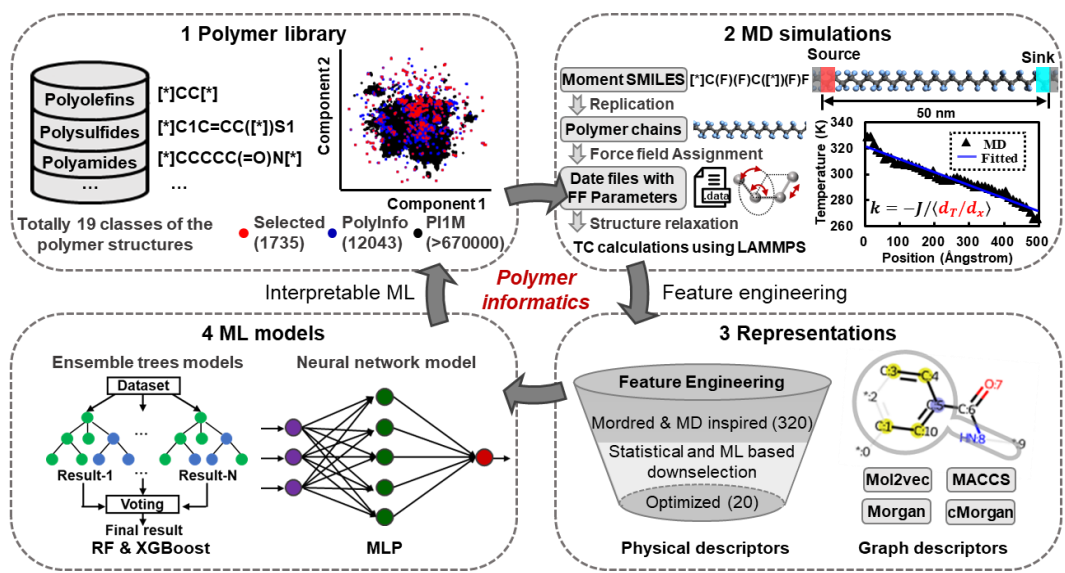

**Fig. 1.** Schematics of high-throughput screening of polymers with high TC via interpretable machine learning.

# 2. RESULTS AND DISCUSSION

## 2.1 Distribution of polymer datasets in chemical space

Polymer data from literatures [51,52] were utilized as the benchmark database for training machine learning models, as well as PolyInfo [53] and PI1M [54] databases were used for the virtual screening of polymer structures with high TC. The polymers are classified into 19 classes such as polyolefins, polyethers and polyamides according to different elements and chemical functional groups [57]. To validate the distribution of the selected 1735 benchmark data over the other two datasets, their chemical structures were visualized in 2D space by the uniform manifold approximation and projection (UAMP) [58], where the chemical structure of each monomer was transformed into the Morgan fingerprint [35] of a 1024 vector with a radius of 2 atoms. It is observed that the polymer structures in the selected benchmark dataset (Fig. 2a) are well covered by the chemical space distribution of those in the PolyInfo (Fig. 2b) and PI1M (Fig. 2c) databases. Note that the PI1M dataset was generated by a generative model of a recurrent neural network trained with data from PolyInfo, which fills the sparse region of the chemical space of the PolyInfo dataset, but the distribution is consistent [54]. Thus, the ML models trained with the selected data are well able to learn the chemical features of all candidates and can be effectively adopted for the virtual screening of polymer structures with high TC.

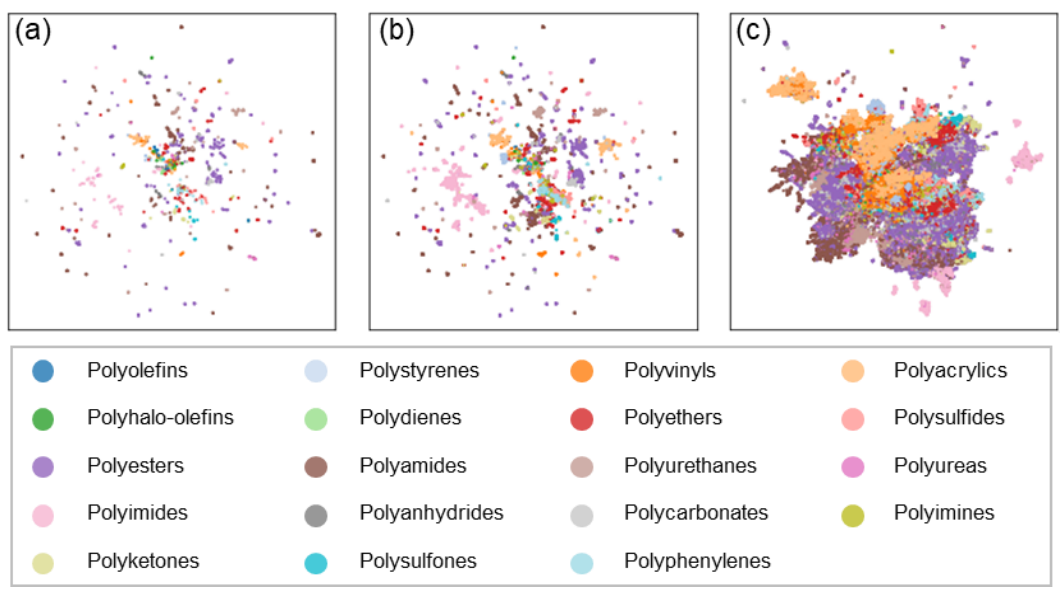

**Fig. 2.** Visualization of polymer data distribution in a 2D space by UMAP. (a), (b) and (c) are corresponding to the selected, PolyInfo and PI1M datasets, respectively.

*2.2 Polymer descriptors hierarchical down-selection and ML Models Training*

Polymer descriptors are hierarchically down-selected in three stages: removing features with low variance, primary filtering referred to different correlation coefficients, and final selection assisted with the ML model (shown in Supplementary Section S1). The collected initial monomer physical descriptors are composed of 286 Mordred-based and 34 MD-inspired descriptors. The descriptors of MD-inspired and Mordred-based descriptors are listed in Supplementary Section S2. The removal of low variance descriptors is intended to eliminate descriptors with variance less than a specific threshold, whose contribution to the target property of all polymer data (TC in this work) is considered to be nearly consistent. After the variance threshold was set to 0.01, the 264 descriptors were reserved for the next stage. We established the weight assignment mechanism (WAM) based on the different correlation coefficients for further primary filtering of the descriptors, due to the various attentions of their mathematical models. The Pearson, Spearman and Distance coefficients are used to evaluate linear, monotonic and non-linear relationships between data respectively, while the maximum information coefficient (MIC) reflects the association of two variables through information entropy, whether linear or nonlinear. The reliability of MIC depends on the data sample size and the value is reliable only with large datasets. The four metric coefficients of Pearson, Spearman, Distance and MIC were incorporated and each was assigned a weighting factor of 0.25, and the thresholds were set to 0.05, 0.05, 0.153 and 0.132, respectively. The 53 descriptors with a cumulative weight value of 1 were retained through VAM. Random sequential feature selection (RFSF) combined with the RF model was then developed for optimized descriptors determination. Considering all possible combinations of descriptors for ML model training is time-consuming and expensive, so traditional SFS usually leads to sub-optimal solutions, where the recommended ensemble of optimized descriptors is not unique, and is influenced by the input order of the descriptors [59]. Here, we disrupted the order of the input descriptors, combined them with 100 RF model training cycles, and acquired the final optimized descriptors based on a statistical approach. The threshold was set to 0.34, that is to maintain descriptors that occur more than 34 times in 100 RF model training runs. The results of the optimized descriptors based on VAM and RSFS are shown in Fig. 3a and their detailed descriptions are listed in Supplementary Section S3. Moreover, Fig. 3e exhibits the Pearson correlation matrices of the correlations among optimized descriptors (Other metrics see in Fig. S1). It is found that most descriptors are positive correlated with each other and negative correlated with TC. Only three descriptors are positive for TC, two of which are MD-inspired descriptors. For example, the descriptor MW_ratio reflects the ratio of the molecular weight of the mainchain to the molecular weight of the monomer, with values between 0 and 1. The MW_ratio of 1 indicates that the polymer is without side chains, which reduces the loss of heat flux along the chain and makes it possible to get large TC.

Figure 3b shows the results of the RF model trained with the optimized descriptors, with training and test $R^2$ of 0.87 and 0.84 respectively. To verify the extensibility of the optimized descriptors, XGBoost and MLP models were deployed for training (see Fig. S2). The accuracy of the training and test sets for XGBoost is 0.95 and 0.87, and that for MLP are 0.81 and 0.88 respectively, which is comparable or even better than the RF model. Therefore, these three models are utilized in the subsequent discussion.

The prediction accuracy of ML models at different down-selection stages illustrated in Figure 3c (training and test data set prediction in Fig. S3). The extra PCA more than 95% variance was performed to compare with RFSF technology. According to the relationship between the number of principal components and the cumulative variance in Fig. S4, at least 19 components are required to exceed 95% variance. This is close to the number of sets of optimized descriptors. As seen in Fig. 3c, the tree-based models of the RF and XGBoost maintain relatively high accuracy even with large descriptor dimensions because of their strong ability to prevent overfitting of the data. Moreover, the feature down-selection process is usually accompanied by the loss of information, which results in the decrease of model accuracy. However, the feature down-selection process also reduces the redundancy between data which suppresses the overfitting and improves the accuracy of the MLP model. Overall, the accuracy of all three models trained with the optimized descriptors from RFSF is higher than that of the models trained with the PCA-derived descriptors, which demonstrates the effectiveness of our approach.

The ML models with different graph descriptors were applied for comparison in Fig. 3d (training and test data set prediction in Fig. S5). The Mol2vec [34] is an unsupervised ML approach to learn vector representations of molecular substructures, which requires a benchmark dataset for molecular structure training. Here, the pre-trained polymer embedding model was from elsewhere [54], which was created using the PolyInfo and PI1M datasets. The MACCS [36] descriptor is the structural key-based descriptor with 166-bit keyset. The Morgan and Morgan count (cMorgan) [35] descriptors are the extended connectivity fingerprints that capture molecular features relevant to molecular activity. The results in Fig. 3d reflect the superiority of ML models trained with the optimized descriptor, no matter the models of RF, XGBoost and MLP. The down-selection processes of physical descriptors examine individual/combined descriptors in relation to TC, while the graph descriptors aim to represent molecular/monomeric information as completely as possible. Whilst the elements or groups in the molecular graph have been indicated to correlate with the TC of polymer chains [30], it is more intuitive and effective to predict the TC of polymer chains using the associated physical descriptors. But not absolute, which is also related to the parameters such as chain stiffness [60].

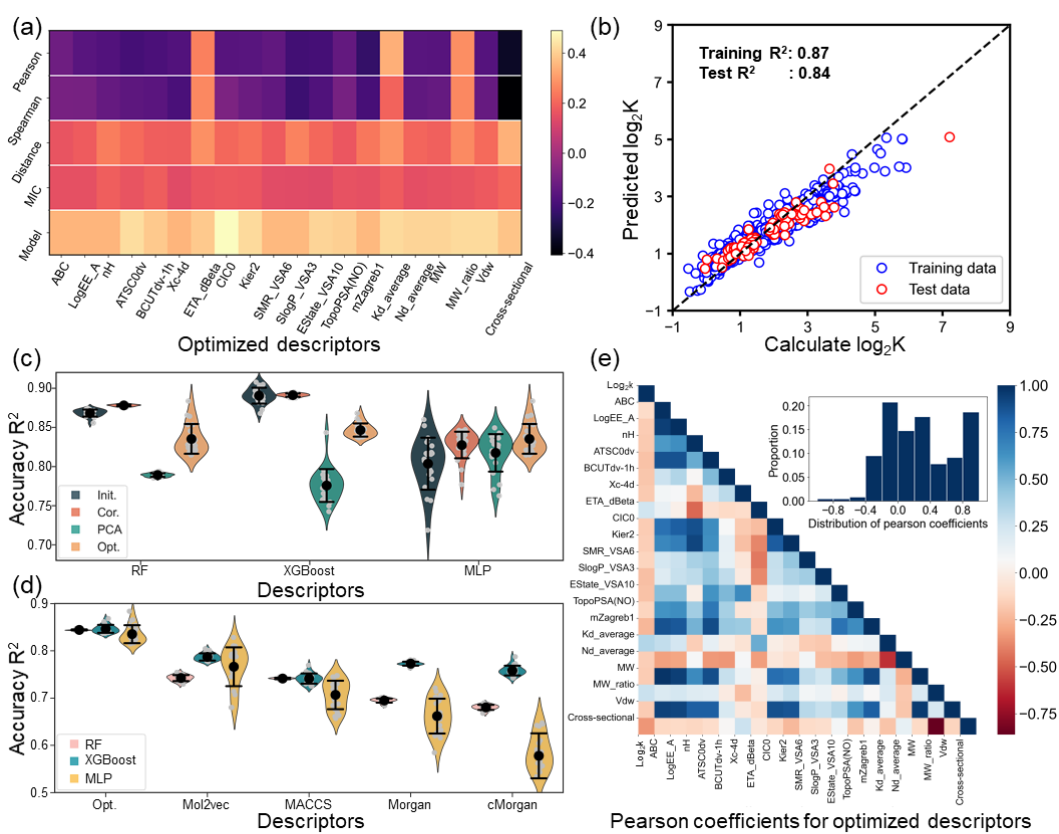

**Fig. 3.** Polymer descriptors down-selection and ML models training. (a) Optimized descriptors acquired by down-selection with four coefficients - Pearson, Spearman, Distance, and MIC coefficients - and RF model. (b) Accuracy of RF model based on optimized descriptors, where training $R^2$ is 0.875 and test $R^2$ is 0.844. (c) Accuracy of ML models at different down-selection processes, including initial (Init.), mathematical correlation (Cor.) coefficients screening and RF model optimization (Opt.) stages. And, an additional PCA approach was applied to compare. (d) Accuracy of ML models with different polymer representation approaches. The violin plot represents the distribution of values, individual subsamples are shown in gray, and the mean and standard of $R^2$ in black. (e) Pearson correlation matrices showing correlations among optimized descriptors and TC. The inset is the statistics of the Pearson coefficients distribution.

## 2.3 Physical insights from interpretable ML model

Figure 4 summarizes the effect of the features using SHAP, for the RF model trained on optimized descriptors. The SHAP approach attempts to address the unexplainable black-box challenge of ML algorithms by calculating the marginal contribution of features to the model output [56]. Hence, the features of each polymer structure in training data sets are assigned the SHAP values separately. As shown in Fig. 4a, the importance ranking of the optimized descriptors was referenced to the average SHAP value. Among the top 8 optimized descriptors, the number of MD-inspired and Mordred-based descriptors is equal, which reflects the fact that the construction of the RF model is a joint contribution of these two types of descriptors. The distribution of SHAP values for each descriptor is displayed in Fig.4b, and the depth of shade of datapoints in the beeswarm plot represents the magnitude of TC of polymer structures in the training set. The distribution of SHAP values for the top-ranked features is relatively wide, and is monotonic about the feature values overall (Fig. S6).

Here, we highlight the two MD-inspired descriptors of cross-sectional and Kd_average. The most important descriptor of cross-sectional indicates the effective cross-sectional area of polymer chain, which is intuitive in relation to the TC. From the Fig. 4c, the SHAP value for cross-sectional decreases monotonically with the descriptor. According to the Fourier's law, the heat flow rate along 1-D polymer chain can be expressed as $Q/d_t = -kA\,d_T/d_x$, where $Q$ is the heat flow, $d_t$ is the time interval, $k$ is the thermal conductivity, $A$ is the cross-sectional area, and $d_T$ and $d_x$ are the temperature difference and distance between the hot and cold ends, respectively [61]. Therefore, the TC is negatively related to the cross-sectional area, and polymers with high TC usually have a small cross-sectional area (Fig.S7a). Moreover, the polymer chain structure is absent of disorder compared to the amorphous structure, maintaining the symmetry of the crystal and reducing phonon scattering. However, the polymer chains may rotate and become disordered due to temperature and other effects, resulting in a rapid decrease in TC [62]. The closely correlation between dihedral angle energy constant and polymer chain stiffness has been demonstrated, and the dihedral angle force constant Kd has been artificially increased in MD simulations to maintain PE chain stiffness and increase thermal conductivity [62,63]. The Kd_average is the average of all types of dihedral force constants from GAFF2 force field for polymer chain, which is roughly proportional to the corresponding SHAP value in Fig. 4d. Especially for polymer structures with great kd_average (>4 kcal/mol) usually have large SHAP values and TC (Fig S7 b). Notably, the TC of polymer chains is influenced by multiple parameters and it is difficult to have the individual descriptor to determine its value. One example is that crystalline polynorbornene has been proved to be weakly sensitive with the chain stiffness, even if increasing the dihedral angular force constant term in MD simulations [62]. This confirms the significance of our proposed ML framework for predicting the TC of polymers.

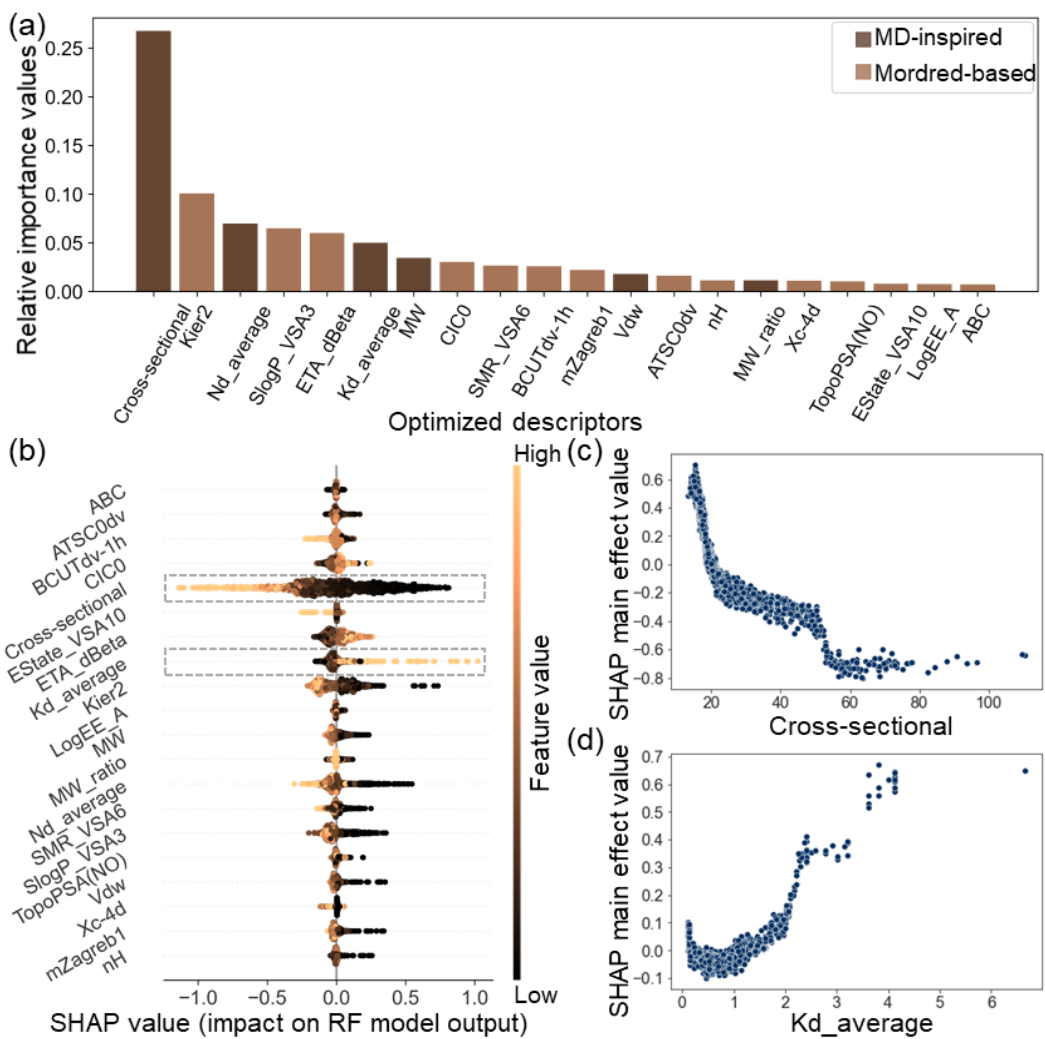

**Fig. 4.** Analysis of feature importance using SHAP on RF model trained by optimized descriptors. (a) Average SHAP values for 20 optimized descriptors. (b) Represent the SHAP values of each descriptor related to training data set polymers in a beeswarm diagram. (c) and (d) SHAP values for the Cross-section and Kd_average of the training data set polymers as a function of descriptor value. The Cross-section is the effective cross-sectional area of polymer chain, and the Kd_average is the average value of force constants of the dihedral angle from GAFF2 force field.

## 2.4 Discovery of high TC polymers

The optimized descriptors were validated reliability in combination with different ML models for predicting the TC of polymer chains. Next, we applied these ML models to predict the TC of polymer structures in the PolyInfo and PI1M databases, in order to virtually screen promising polymers with high TC. The predicted polymer TC versus cross-sectional area from the ensemble of optimized descriptors combined with RF, XGBoost and MLP are visualized in Fig. 5a-c, respectively. Where stars indicate polyethylene with $\log_2$TC of 3.91, 4.66, and 5.30 predicted by RF, XGBoost, and MLP respectively, and that calculated by MD simulation is 5.28. The dependence of TC on the cross-sectional area is evident here, as almost all of the predicted high TC polymers have small cross-sectional areas. Moreover, since PI1M has the same chemical distribution space as PolyInfo and fills the sparse area, which covers most of the TC range of PolyInfo and enriches the polymer structures in the high TC region.

Comparing the results from different ML models, the tree-based models of RF and XGBoost predict the TC of polymers in a narrower space than that of the MLP. Though the excellent performance of the tree-based models in preventing overfitting, the extrapolation of the models is usually inadequate and the predictions are still limited to the range of TC of the polymer structures in the training set. In contrast, neural network model of MLP usually has better extrapolation capability, and is superior in finding small data such as high TC polymer structures, despite the relatively low training accuracy $R^2$ of the model. This finding is similar to previous study of predicting the permeability of gas separation membranes using ML [17]. As well, previous work has revealed the length dependence of the thermal conductivity of polymer chains. Within a certain length range, the diverging thermal conductivity $k$ and chain length $L$ can be fitted by $k \sim L^{\beta}$, where $\beta$ indicates the relatively dominant phonon transport mechanism [13]. Here, we considered polymer chains with TC greater than 20.00 W/mK with an effective length of 50 nm as the outstanding polymers with high TC. Then, balanced strategy to integrate the three ML models performance were devised to recommend promising polymer structures for calculation of TC by MD simulations. We identified the polymer structures in the PolyInfo dataset with RF, XGBoost and MLP predictions of $\log_2$TC up to 3.51, 3.50 and 4.33, and only the polymer structures with no less than 2 occurrences were picked for MD simulations. As a result, 24 polymer structures with high TC were discovered and verified. Similarly, we implemented this method to identify 84 high TC polymer structures in the PI1M database. After de-duplication, totally 107 high TC polymer structures were found in this work, and the Synthetic accessibility (SA) scores were calculated as shown in Fig. 5d. The specific polymer structures can be seen in Supplementary Section S8. From Figure S8, we can see that most of the high TC polymers are simple linear or contain aromatic rings in the mainchain, which have small repeating unit lengths and no side chains. The SA score was initially utilized to estimate the synthetic accessibility of drug-like molecules based on molecular complexity and fragment contributions [64], and was subsequently adopted for polymers [31,32]. The SA score values ranged from 1 to 10, and synthesis is more difficult as the value increases. To take into account the effect of monomer linkages, polymer molecules with a polymerization degree of 6 were calculated for the SA score. Among them, 28 polymer structures with SA no more than 3.00, including polyethylene, polytetrafluoroethylene and poly(p-phenylene), and etc. Although it is currently difficult to fabricate each of these structures, we believe that more polymers like PE chain will be prepared for exploring the limits TC of polymers by combining advanced processes such as micromechanical stretching, electrostatic spinning and nanotemplate preparation in the near future [4,6,11].

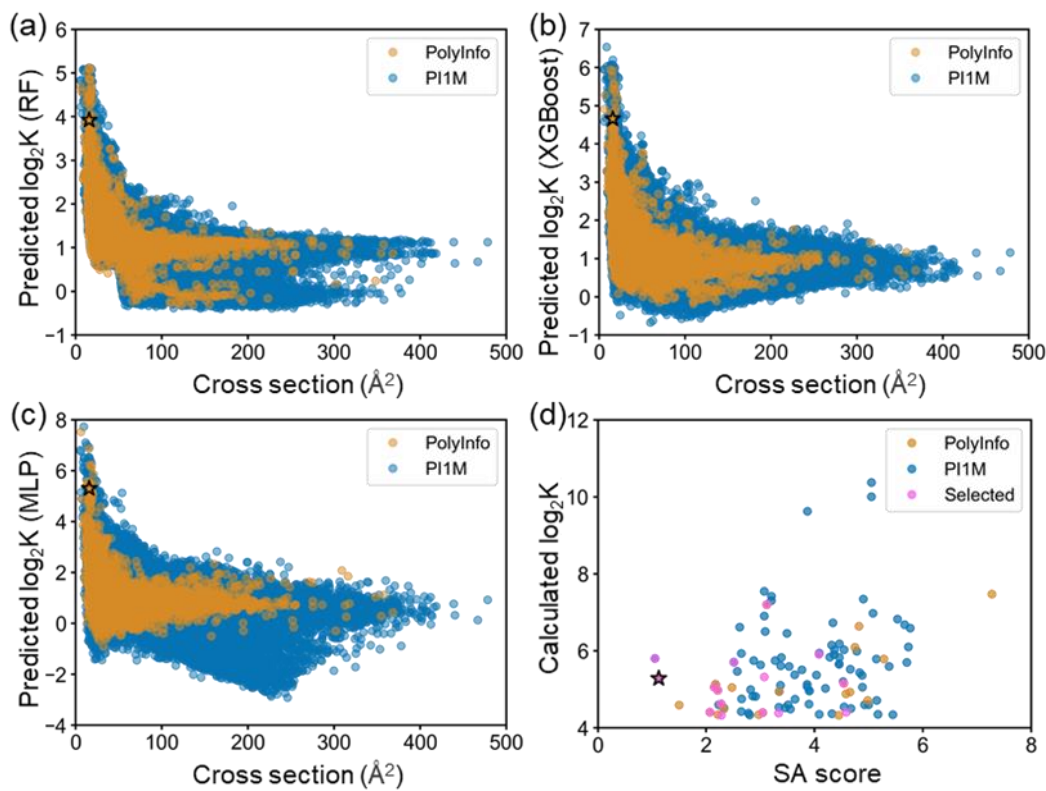

**Fig. 5.** Prediction of high TC polymers in PolyInfo and PI1M databases using constructed ML models. (a), (b) and (c) based on RF, XGBoost and MLP models, respectively. (d) Synthetic accessibility (SA) score versus calculated $\log_2 K$ of screened high TC polymers (TC > 20.00 W/mK). The star indicates PE, and the TC in this work is 38.98 W/mK.

*2.5 Symbolic regression for prediction of promising polymers*

Since the TC of polymer chains is influenced by complex multi-parameters, it is difficult to predict trends in TC values for different polymers from any single descriptor. Symbolic regression (SR) attempts to accelerate the discovery of materials with superior properties by relating available descriptors through mathematical formulas to construct new combinatorial features [65]. SR does not require massive datasets, as long as a high consistency and accuracy [66,67]. The 107 promising polymer structures (TC > 20.00 W/mK) with optimized descriptors were utilized for SR, where the ratio of training to test set was 3:1. The mathematical formula were acquired and selected using an efficient stepwise strategy with SR based on genetic programming (GPSR) as implemented in the gplearn code [68]. The hyperparameters setup and the detailed formula determination process can be found in Supplementary Section S9. Pearson coefficients are first applied to filter optimized descriptors and create sub-descriptors, and a novel ensemble of 22 descriptors was obtained. The frequency of occurrence of optimized descriptors in 158 mathematical formulas (PC values >=0.85 and complexity <=10) is displayed in Fig. 6a, and first 8 descriptors were finally retained. It is worth emphasizing that the MD-inspired descriptors of cross-sectional area (cross-sectional) and dihedral force constants (Kd_average) appeared in each of the formulas. In Figure 6b, we calculated the Pearson coefficients of the new set of descriptors with the TC, the results suggest these descriptors are closely associated with the TC. Subsequently, we reset the grid search hyperparameters in gplearn and used $R^2$ as the evaluation metric. Only formulas with high $R^2$ and low complexity (length of formula) are considered suitable for the prediction the TC of polymer structures [69]. Thus, 9073 mathematical formulas with complexity within 30 and $R^2$ over 0.6, which are characterized by complexity and accuracy $R^2$ via density plot in Fig. 6c. The four points of c, d, e and f at Pareto front were identified by Latin hypercube sampling approach [70,71], and their corresponding formulas are expressed in Table S8. The complexities of the four formulas are in the range of 20 to 30, and the fitting accuracies are all greater than 0.70. Moreover, the training accuracy is mostly positive to complexity. For example, the formula represented by point c with complexity of 20 has a relatively low accuracy $R^2$ among the four points, but the fitting results are consistent with the MD labeled $log_2K$, as demonstrated in Fig. 6d. Meanwhile, all four identified formulas include the descriptors of the Cross-sectional, Kd_average and Nd_average, which verified that the TC of polymer chain is strongly correlated with the parameters such as cross-sectional area and dihedral stiffness. These formulas are meaningful in the initial rapid screening of high TC polymer chain structures.

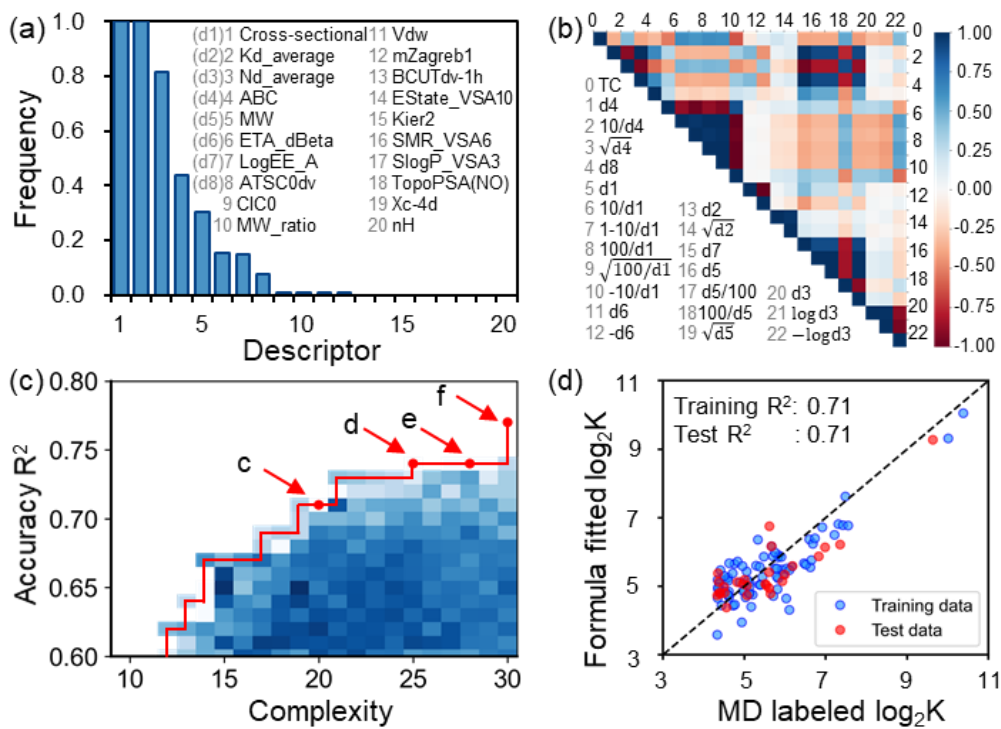

**Fig. 6.** GPSR for TC prediction of promising polymers. (a) Frequency of occurrence of optimized descriptors in 158 mathematical formulas (PC values >=0.85 and complexity <=10). (b) Pearson correlation matrices showing correlations among 22 descriptors and TC, where the descriptors d1~d8 correspond to descriptors 1 to 8 in Fig. 6a. (c) Pareto front of accuracy $R^2$ vs. complexity of 9073 mathematical formulas shown via density plot. (d) MD labeled vs. fitting results of the formula (point c) with complexity of 20 and training accuracy $R^2$ of 0.71.

*2.6 Thermal transport mechanism of promising crystal polymers*

Taking into account factors such as TC and SA score, eight polymer structures (see in Fig. 7a) were chosen for the analysis of phonon dispersion relations. Currently, polymer structures like [*]C=C[*] and [*]N=N[*] are challenging to be synthesized experimentally, but are contributing to our understanding of polymer thermal conductivity mechanisms. All of these polymer molecules are π-conjugated structures except for the PE and the Polytetrafluoroethylene (PTFE), which are simple linear structures. In π-conjugated polymer molecules, the overlap of p-orbitals has enhanced restraint in inhibiting chain rotation and forming the rigid backbone [12]. Figure 7b illustrates the phonon dispersion relations, which were obtained by phonon spectral energy density (Phonon-SED) analysis [72], The detailed description of the Phonon-SED approach can be found in the Method part. Since the acoustic modes are dominated by the thermal transport of heat carriers in polymer crystals, phonon modes with frequencies below 25 THz are demonstrated. Moreover, the phonon group velocity is approximated as the average of the slopes of all acoustic branches [12,60]. In the one-dimensional polymer chain systems, the TC is analyzed as $k = v_g C_v l$, where $v_g$ is the phonon group velocity, $C_v$ is the volumetric heat capacity and $l$ is the phonon mean free path. Due to the limitation of classical MD simulations, the volumetric heat capacity can be expressed as $C_v = 3k_b N/V$, where $k_b$ is Boltzmann constant, $N$ is the number of atoms and $V$ is the volume. Thus, phonon mean free path can be calculated by the ratio of the TC to the multiplication of the phonon group velocity and the volumetric heat capacity. The approximations of the above calculations allow the results to be rough, but it do help us to understand the underlying thermal conductivity mechanisms of these promising polymer structures by comparing the relative trends of the relevant parameters, as listed in Table 1.

The volumetric heat capacity of the eight polymer structures varies from 3.28 to 6.13, which is not critical to the high TC of polymer chains. As for the phonon group velocity, the six π-conjugated polymers have large values (more than 5900 m/s) due to overlapped p-orbital and delocalized electrons. Additionally, the small atomic mass enables a large phonon group velocity. The PTFE has smaller phonon group velocity than that of PE due to the relatively larger mass of fluorine atoms compared to hydrogen atoms. The phonon mean free path provides valuable insights into phonon transport in the polymer chains. Overall, simple linear polymer chains are easily to have long phonon mean free paths, especially for linear π-conjugated polymers of [*]C=C[*] and [*]N=N[*]. These structures have large chain stiffness and few atoms except for the backbone, thereby having weak phonon-disorder scattering.

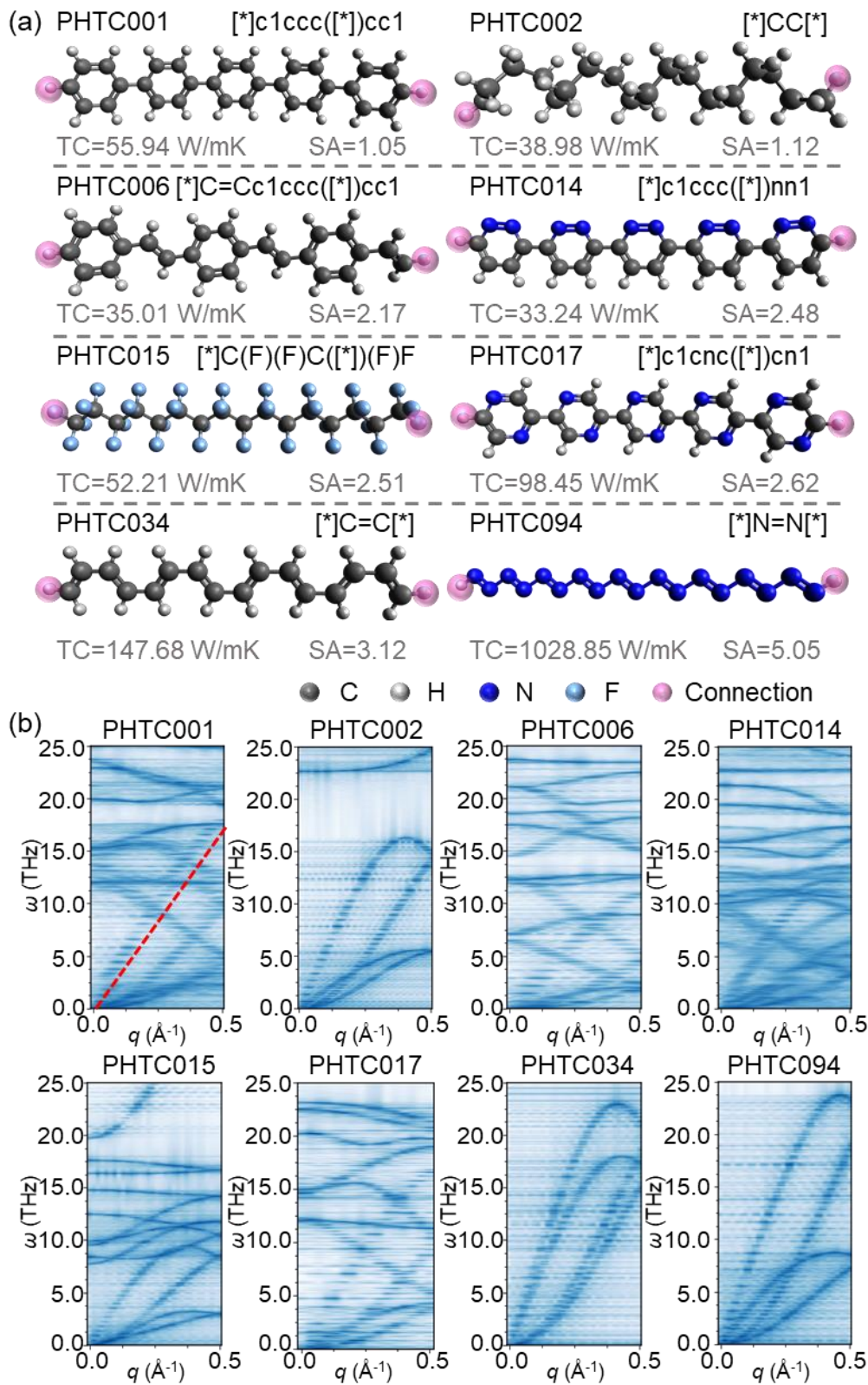

**Fig. 7.** Structure and phonon dispersion relations for the eight promising polymers. (a) Polymer chain structures. (b) Phonon dispersion relations. The $q$ is wavevector, the $\omega$ is phonon frequency and the average phonon group velocity of one branch is estimated as the slope of the origin to the maximum frequency point as shown in the red dashed line in the PHTC001 structure.

**Table 1.** Volumetric heat capacity, phonon group velocity, phonon mean free path and phonon thermal conductivity for the eight promising polymers

| Polymer ID | SMILES | SA | $C_v$ (J/cm$^3$K) | $v_g$ (m/s) | $l$ (nm) | $k$ (W/mK) |
|---|---|---|---|---|---|---|
| PHTC001 | [*]c1ccc([*])cc1 | 1.05 | 5.27 | 6822.21 | 1.55 | 55.94 |
| PHTC002 | [*]CC[*] | 1.12 | 6.13 | 5240.91 | 1.21 | 38.98 |
| PHTC006 | [*]C=Cc1ccc([*])cc1 | 2.17 | 5.25 | 6295.54 | 1.06 | 35.01 |
| PHTC014 | [*]c1ccc([*])nn1 | 2.48 | 5.03 | 5927.05 | 1.12 | 33.24 |
| PHTC015 | [*]C(F)(F)C([*])(F)F | 2.51 | 3.84 | 2952.11 | 4.61 | 52.21 |
| PHTC017 | [*]c1cnc([*])cn1 | 2.62 | 5.03 | 7439.50 | 2.63 | 98.45 |
| PHTC034 | [*]C=C[*] | 3.12 | 4.37 | 8380.09 | 4.03 | 147.68 |
| PHTC094 | [*]N=N[*] | 5.05 | 3.28 | 6378.73 | 49.22 | 1028.85 |

## 3. CONCLUSIONS

We have developed an interpretable ML framework for exploring high thermal conductivity polymer chains via high-throughput MD simulations. Inspired by the drug-like small molecule representation and the molecular force field, we reduced the initially calculated/collected 320 physical descriptors to 20 optimized descriptors by hierarchical down-selection. The constructed ML models are capable of effectively reflecting the relationship between optimized descriptors and property, and exhibit high accuracy in TC prediction. All the models of RF, XGBoost and MLP achieved the $R^2$ of more than 0.80, which is superior to that of represented by conventional graph descriptors. Moreover, the promotion or inhibition of TC by optimized descriptors like cross-sectional area and dihedral stiffness was captured by RF model using SHAP analysis.

Using the trained ML models, we discovered 107 promising polymers with TC greater than 20.00 W/mK, and 29 of which have SA scores no more than 3.00. These polymer structures have been validated through high-fidelity MD simulations. Further, we used SR with optimized descriptors to fit the TC of promising polymers, and the derived mathematical formulas enable a preliminary fast screening of high TC polymers without relying on ML models, which is friendly for experimental studies. In closing, we calculated phonon dispersion relations for eight typical polymer structures via phonon spectral energy density analysis to reveal the underlying TC mechanisms. Notably, most of these structures are π-conjugated polymers, whose overlapping p-orbitals enable easily to maintain strong chain stiffness and large group velocities. Our approach may assist in the research of high-performance polymers that are not limited to TC.

# 4. METHODS

*4.1 Polymer modeling and cross-sectional area calculation*

Polymer modeling is a monomer to chain process, implemented in the STK tool, with input parameters of monomer SMILES and degree of polymerization [73]. The length of the polymer chains was set uniformly to 50 nm, and the degree of polymerization was obtained by dividing the chain length by the monomer length and rounding up to an integer. Starting from the polymer SMILES, a molecular chain with polymerization degree 2 was generated by RDKit and optimized using the MMFF force field [74]. Then, the monomer length was determined by measuring the distance between equivalent atoms in two repeating units in the heat transport direction. Following the modeling, a Python pipeline of PYSIMM realized the assignment of GAFF2 force field parameters and the generation of MD simulation input structure data files [75].

The cross-sectional area is one of the important parameters for thermal conductivity analysis. In molecular dynamics simulations, the calculation of the cross-sectional area is difficult for systems that do not occupy the entire simulation box. The cross-sectional area was estimated by the ratio of the van der Waals volume to the length of the monomer [9]. The Van der Waals volume of the monomer was calculated by the sum of atomic and bond contributions, and has been successfully tested and applied in previous drug compounds [76].

*4.2 Calculation of TC by MD simulations*

The TC of Polymer chains were obtained by non-equilibrium molecular dynamics simulations (NEMD) performed in a Large-scale Atomic/Molecular Massively Parallel Simulator (LAMMPS) [75]. In terms of the NEMD method, the heat energy exchange was achieved by an enhanced version of the heat exchange algorithm, which rescales and shifts the velocities of particles inside reservoirs to impose a constant heat flux [77]. The polymer chains were placed in a box of 540×60×60 ($x$×$y$×$z$) Å box, where the dimension in the $y$ and $z$ directions was set to 60 Å to avoid interaction with the neighboring polymer chains. Before TC calculation, the polymer chain structures were relaxed to reach a stable conformation. Then, the polymer chain was divided into 50 slabs in the $x$ direction, and the fixed regions at two ends of the chain were set as a heat-insulating walls. In the NEMD simulation, the system was run under NVT (constant number of atoms, volume, and temperature) and NVE (constant number of atoms, volume, and energy) ensembles for 1 ns at 300 K sequentially to release chain stress [30,78]. After that, the heat was added/extracted to the heat source/sink regions (20 Å of each region) at the end of the polymer chain in a regular rate to create a constant heat flux. The applied heat varies for different polymer chain structures and ranged from 0.01 eV/ps to 0.08 eV/ps. At last, the temperature profile was averaged over the last 2~3 ns and used for TC calculation based on Fourier's law $k = -J(d_T/d_x)$, where $J$ is heat flux, $d_T/d_x$ is the temperature gradient.

### 4.3 Descriptors calculation and ML models construction

The ideal polymer descriptors are required to minimize and completely represent polymer information, and is one of the key factors in determining the prediction accuracy of ML algorithms. The physical descriptors for this work were sourced from both Mordred software calculations and GAFF2 force field parameters extraction. The Mordred software was initially developed for small molecule characteristics in cheminformatics, which can calculate more than 1800 descriptors [41]. However, since we consider two linkages of polymer monomers, only 286 valid descriptors were obtained. Therefore, as a complement, we additionally extracted parameters from each polymer force field file as the descriptor. For graph descriptors, MACCS, Morgan and cMorgan fingerprints were calculated in the RDKit package [39]. The Mol2vec fingerprints were embedded via Mol2vec [34]. We referred the polymer representation model trained using PoLyInfo and PI1M databases [54].

The ML models of RF, XGBoost and MLP were implemented by using Scikit-learn [79]. Hyperparametric optimization for RF, XGBoost and MLP was operated with the Bayesian Optimization package [80] which is a global optimization tool to achieve good prediction accuracy $R^2$. The Gaussian regression process and acquisition function with 10 randomly pairs of parameters were selected for initial training, and the ideal parameters for each ML model were determined after 100 optimization iterations [46].

To explain the association of optimized descriptors with TC, we used the SHAP toolkit with RF model to evaluate the feature importance [56]. The SHAP analysis is based on a game-theoretic approach that associates the optimal credit allocation with the local explanations of the model, which considers the model performance by neglecting each feature and provides direction of each descriptor effect [46].

### 4.4 Mathematical formulas for TC fitted by symbolic regression (SR)

The mathematical formulae were acquired and selected using an efficient stepwise strategy with SR based on genetic programming (GPSR) as implemented in gplearn [68]. The 107 polymer structures with TC greater than 20.00 W/mK were randomly divided into 3:1 as training and test sets respectively. At first, Pearson coefficients were used as evaluation metrics of training fitness to filter optimized descriptors and generate sub-descriptors, and a new dataset containing 22 descriptors was generated. Further, the grid search strategy with the hyperparameters and metric $R^2$ as listed in Table 2 was applied to determine the mathematical formulas. We ultimately discussed four formulas at Pareto front were identified by Latin hypercube sampling approach [70,71]. More information about SR can be found in the Supplementary Section S9.

**Table 2.** Setup of hyperparameters in gplearn for GPSR

| Parameter | Value |
| --- | --- |
| Generations | 300 |
| Population size in every generation | 5000 |
| Probability of crossover (pc) | [0.30,0.90], step=0.05 |
| Subtree mutation (ps) | [(1-pc)/3,(1-pc)/2] (step= 0.01) |
| Hoist mutation (ph) | [(1-pc)/3,(1-pc)/2] (step = 0.01) |
| Point mutation (pp) | 1-pc-ps-ph |
| Function set | $\{+, -, \times, \div, \sqrt{x}, \ln x, |x|, -x, 1/x\}$ |
| Parsimony coefficient | 0.001, 0.003, 0.005 |
| Metric | $R^2$ |
| Stopping criterial | 0.900 |
| Random_state | 0, 1, 2, 3, 4 |
| Init_depth | [2, 6], [4, 8], [6, 10], [2, 10] |

*4.5 Analysis of phonon dispersion relations by phonon spectral energy density (Phonon-SED)*

To understand the TC mechanism of polymers, MD simulations coupled with Phonon-SED approach [72] were employed to calculate the dispersion relations of polymers. The polymer chain with a length of 100 Å was constructed as an input of SMILES and placed into a box with the cross section of 60×60 Å. After energy minimization, the system was run under the NVT (constant number of atoms, volume, and temperature) ensemble for 0.25 ns at 2 K sequentially to release chain stress. Subsequently, the system was run under the NVE (constant number of atoms, volume, and energy) ensemble for 2 million steps with the timestep of 0.25 fs. During this period, the velocity and position of each atom in the polymer backbone were recorded with intervals of 20 steps. The Phonon-SED converted the time domain information of atomic velocities and positions into wave vectors versus angular frequencies via two-dimensional Fourier transform, expressed as

$$\Phi(q, \omega) = \frac{1}{4\pi\tau_0 N_T} \sum_{\alpha}^{\{x,y,z\}} \sum_{b}^{B} m_b \left| \int_0^{\tau_0} \sum_n^{N_T} \dot{u}_\alpha(n, b; t) \times e^{iq \cdot r(n,0;t) - i\omega t} dt \right|^2 \qquad (1)$$

Where $q$ is the wavevector, $\omega$ is the frequency, $\tau_0$ is the simulation time, $m_b$ is the mass of atom $b$, $\alpha$ is the cartesian direction, $N_T$ is the number of the unit cell in the polymer chain, $\dot{u}_\alpha(n, b; t)$ is the velocity of atom $b$ in the unit cell $n$ at time $t$ in the $\alpha$ direction, and $r(n, 0; t)$ is the equilibrium position of unit cell $n$.

## Acknowledgements

This work was supported by Shanghai Pujiang Program (No. 20PJ1407500), the National Natural Science Foundation of China (No. 52006134), Shanghai Key Fundamental Research Grant (No. 21JC1403300), SJTU-Warwick Joint Seed Fund. The computations in this paper were run on the $\pi$ 2.0 cluster supported by the Center for High Performance Computing at Shanghai Jiao Tong University.

## Author Contributions

Conceptualization: S. J.

Methodology: X. H., S. M. and S. J.

Investigation: X. H., S. M. and S. J.

Supervision: S. J., C. Y. Z. and H. W.

Writing—original draft: X. H., S. M. and S. J.

Writing—review & editing: X. H., S. M., S. J., C. Y. Z. and H. W.

## Competing interests

Authors declare that they have no competing interests.

## Supporting Information

The supporting information is available free of charge.

Supplemental Materials for

# Exploring High Thermal Conductivity Polymers via Interpretable Machine Learning with Physical Descriptors

**Xiang Huang[1,†], Shengluo Ma[1,†], C. Y. Zhao[1], Hong Wang[2], and Shenghong Ju[1, 2, *]**

[1] China-UK Low Carbon College, Shanghai Jiao Tong University, Shanghai, China

[2] Materials Genome Initiative Center, School of Material Science and Engineering, Shanghai Jiao Tong University, Shanghai, China

†Equal contribution

---
* Corresponding author: shenghong.ju@sjtu.edu.cn.

## Section S1. Downselection of polymer descriptors

Polymer descriptors were downselected in three stages: removing features with low variance (Var.), primary filtering referred to different correlation coefficients (Cor.), and final selection assisted with ML model (Opt.), as shown in Fig. S1.

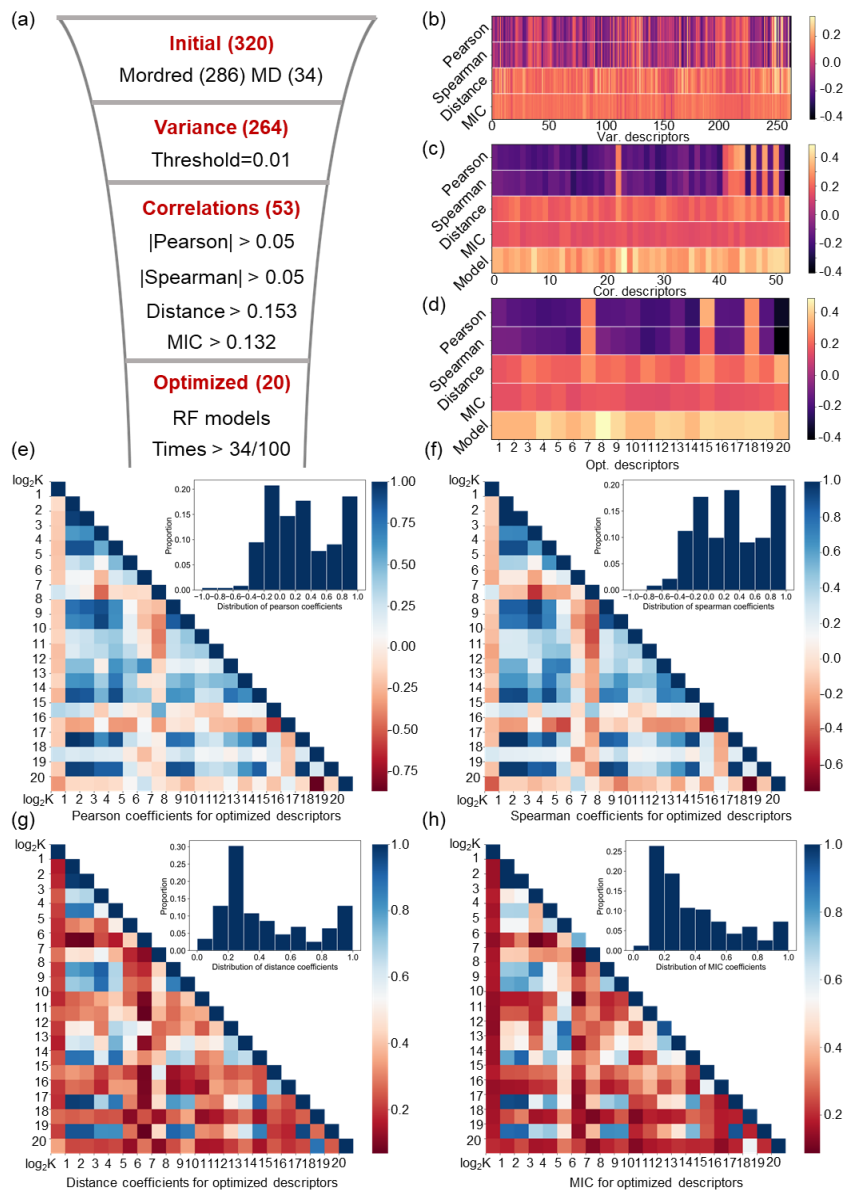

**Fig. S1.** Downselection and analysis of descriptors. (a) Process of descriptors downselection. (b), (c) and (d) Relationship between descriptors at different stages (Var., Cor. and Opt.) and thermal conductivity based on various metrics or frequency of occurrence of descriptors though 100 Random Forest (RF) model selections. (e), (f), (g) and (h) Pearson, Spearman, Distance and MIC correlation matrices showing correlations among optimized descriptors and TC. The inset shows the distribution statistics of the corresponding correlation coefficient values respectively. The description of the 20 optimized descriptors is available in Section S3.

**Section S2. List of MD-inspired and Mordred-based fingerprints**

The MD-inspired descriptors include force field related (FF-related) descriptors and thermal conductivity related (TC- related) descriptors, listed in Table S1.

**Table S1.** Descriptions of all MD-inspired descriptors

| Name | Description |
|---|---|
| Epsilon_max | Maximum value of the depth of the energy potential (Lennard–Jones parameter) |
| Epsilon_min | Minimum value of the depth of the energy potential (Lennard–Jones parameter) |
| Epsilon_average | Average value of the depth of the energy potential (Lennard–Jones parameter) |
| Sigma_max | Maximum value of the equilibrium distance (Lennard–Jones parameter) |
| Sigma_min | Minimum value of the equilibrium distance (Lennard–Jones parameter) |
| Sigma_average | Average value of the equilibrium distance (Lennard–Jones parameter) |
| Kb_max | Maximum value of force constants of the bond |
| Kb_min | Minimum value of force constants of the bond |
| Kb_average | Average value of force constants of the bond |
| R0_max | Maximum value of equilibration structural parameters of the bond |
| R0_min | Minimum value of equilibration structural parameters of the bond |
| R0_average | Average value of equilibration structural parameters of the bond |
| Ka_max | Maximum value of force constants of the bond angle |
| Ka_min | Minimum value of force constants of the bond angle |
| Ka_average | Average value of force constants of the bond angle |
| Theta0_max | Maximum value of equilibration structural parameters of the bond angle |
| Theta0_min | Minimum value of equilibration structural parameters of the bond angle |
| Theta0_average | Average value of equilibration structural parameters of the bond angle |
| Kd_max | Maximum value of force constants of the dihedral angle |
| Kd_min | Minimum value of force constants of the dihedral angle |
| Kd_average | Average value of force constants of the dihedral angle |
| Nd_max | Maximum value of multiplicity for the torsional angle parameters |
| Nd_min | Minimum value of multiplicity for the torsional angle parameters |
| Nd_average | Average value of multiplicity for the torsional angle parameters |
| Delta_max | Maximum value of phase angle for the torsional angle parameters |
| Delta_min | Minimum value of phase angle for the torsional angle parameters |
| Delta_average | Average value of phase angle for the torsional angle parameters |
| Ki_max | Maximum value of force constants of the improper angle |
| Ki_min | Minimum value of force constants of the improper angle |
| Ki_average | Average value of force constants of the improper angle |
| MW | Molecular weight of monomer |
| MW_ratio | The ratio of the molecular weight of the main chain to that of the monomer |
| Vdw | Van der Waals volume of the monomer |
| Cross-sectional | Equivalent cross-sectional area of polymer chain |

The Mordred-based descriptors were calculated by Mordred software [1], the 286 Mordred-based descriptors in this work are listed in Table S2, and the detailed meaning could be found at https://mordred-descriptor.github.io/documentation/master/descriptors.html (accessed Aug 10, 2022 ).

**Table S2.** Descriptions of 286 Mordred-based descriptors

| ABC | nAcid | nBase | SpMax_A | SpMAD_A |
|---|---|---|---|---|
| LogEE_A | VE1_A | VR1_A | VR3_A | nAromAtom |
| nAtom | nBridgehead | nHetero | nH | nN |
| nO | nS | nF | nCl | nBr |
| ATS0dv | ATS0Z | AATS0dv | AATS0d | AATS2d |
| AATS0Z | AATS1Z | AATS2Z | AATS3Z | ATSC0dv |
| ATSC1dv | ATSC2dv | ATSC3dv | ATSC4dv | ATSC5dv |
| ATSC6dv | ATSC7dv | ATSC8dv | ATSC1d | ATSC2d |
| ATSC3d | ATSC4d | ATSC5d | ATSC6d | ATSC7d |
| ATSC8d | ATSC0Z | ATSC1Z | ATSC2Z | ATSC3Z |
| ATSC4Z | ATSC5Z | ATSC6Z | ATSC7Z | ATSC8Z |
| AATSC0dv | AATSC1dv | AATSC2dv | AATSC3dv | AATSC0d |
| AATSC1d | AATSC2d | AATSC3d | AATSC0Z | AATSC1Z |
| AATSC2Z | AATSC3Z | MATS1dv | MATS2dv | MATS3dv |
| MATS1d | MATS2d | MATS3d | MATS1Z | MATS2Z |
| MATS3Z | GATS1dv | GATS2dv | GATS3dv | GATS1d |
| GATS2d | GATS3d | GATS1Z | GATS2Z | GATS3Z |
| BCUTdv-1h | BCUTdv-1l | BCUTd-1h | BCUTd-1l | BCUTZ-1h |
| BCUTZ-1l | BalabanJ | BertzCT | nBondsD | nBondsT |
| C1SP1 | C2SP1 | C1SP2 | C3SP2 | C1SP3 |
| C2SP3 | C3SP3 | C4SP3 | HybRatio | Xch-4d |
| Xch-5d | Xch-6d | Xch-4dv | Xc-3d | Xc-4d |
| Xc-5d | Xc-6d | Xpc-0d | AXp-0d | AXp-1d |
| AXp-2d | AXp-3d | NsCH3 | NdCH2 | NssCH2 |
| NdsCH | NsssCH | NddC | NdssC | NaaaC |
| NssssC | NsNH2 | NssNH | NaaNH | NdsN |
| NaaN | NsssN | NaasN | NsOH | NdO |
| NssO | NaaO | NdS | NssS | NaaS |
| NddssS | SsssCH | SdssC | SaasC | SaaaC |
| SsssN | SaasN | AETA_beta | AETA_beta_s | ETA_beta_ns_d |
| AETA_beta_ns_d | AETA_eta_RL | AETA_eta_BR | ETA_epsilon_3 | ETA_dBeta |
| fragCpx | fMF | nHBAcc | IC0 | IC1 |
| IC2 | IC3 | SIC0 | SIC1 | SIC2 |
| SIC3 | SIC4 | BIC5 | CIC0 | MIC0 |
| MIC1 | MIC2 | Kier2 | Kier3 | FilterItLogS |
| PEOE_VSA1 | PEOE_VSA2 | PEOE_VSA3 | PEOE_VSA4 | PEOE_VSA6 |
| PEOE_VSA7 | PEOE_VSA8 | PEOE_VSA9 | PEOE_VSA10 | PEOE_VSA11 |
| PEOE_VSA12 | PEOE_VSA13 | SMR_VSA1 | SMR_VSA3 | SMR_VSA4 |
| SMR_VSA6 | SMR_VSA9 | SlogP_VSA1 | SlogP_VSA2 | SlogP_VSA3 |
| SlogP_VSA4 | SlogP_VSA5 | SlogP_VSA7 | SlogP_VSA8 | SlogP_VSA10 |
| SlogP_VSA11 | EState_VSA1 | EState_VSA2 | EState_VSA3 | EState_VSA4 |
| EState_VSA5 | EState_VSA6 | EState_VSA7 | EState_VSA8 | EState_VSA9 |
| EState_VSA10 | VSA_EState1 | VSA_EState2 | VSA_EState3 | VSA_EState4 |
| VSA_EState5 | VSA_EState7 | VSA_EState8 | VSA_EState9 | AMID_h |
| AMID_N | AMID_O | AMID_X | nRing | n5Ring |
| n6Ring | n7Ring | nHRing | n4HRing | n5HRing |
| n6HRing | n5aRing | naHRing | n6aHRing | nARing |
| n5ARing | n6ARing | nAHRing | n5AHRing | n6AHRing |
| nFRing | n7FRing | n8FRing | n9FRing | n10FRing |
| n11FRing | n12FRing | nG12FRing | nFHRing | n7FHRing |
| n8FHRing | n10FHRing | nG12FHRing | nFaRing | n8FaRing |
| n9FaRing | n10FaRing | nG12FaRing | nFaHRing | nFARing |
| n10FARing | n12FARing | nG12FARing | nFAHRing | nG12FAHRing |
| RotRatio | SLogP | TopoPSA(NO) | TopoPSA | GGI4 |
| GGI5 | GGI6 | GGI7 | JGI1 | JGI2 |
| JGI3 | JGI4 | JGI5 | JGI6 | JGI7 |
| JGI8 | JGI9 | JGI10 | JGT10 | TopoShapeIndex |
| mZagreb1 | | | | |

**Section S3. Descriptions of optimized descriptors for polymer thermal conductivity (TC) prediction**

Polymer optimized descriptors were obtained by downselection, listed in Table S3.

**Table S3.** Descriptions of all MD-inspired descriptors

| No. | Name | Description | Block |
|---|---|---|---|
| 1 | ABC | Atom-bond connectivity index | Atom-bond connectivity index descriptor |
| 2 | LogEE_A | LogEE of adjacency matrix | Adjacency matrix descriptor |
| 3 | Nh | Number of H atoms | Atom count descriptor |
| 4 | ATSC0dv | Centered moreau-broto autocorrelation of lag 0 weighted by valence electrons | Centered Autocorrelation of Topological Structure descriptor |
| 5 | BCUTdv-1h | First heighest eigenvalue of Burden matrix weighted by valence electrons | BCUT descriptor |
| 6 | Xc-4d | 4-ordered Chi cluster weighted by sigma electrons | Chi descriptor |
| 7 | ETA_dBeta | ETA delta beta | ETA delta beta descriptor |
| 8 | CIC0 | 0-ordered complementary information content | Complementary information content descriptor |
| 9 | Kier2 | kappa shape index 2 | Kappa shape index 2 descriptor |
| 10 | SMR_VSA6 | MOE MR VSA Descriptor 6 ( 2.75 <= x < 3.05) | MOE type descriptors using Wildman-Crippen MR and surface area contribution |
| 11 | SlogP_VSA3 | MOE logP VSA Descriptor 3 (-0.20 <= x < 0.00) | MOE type descriptors using Wildman-Crippen LogP and surface area contribution |
| 12 | EState_VSA10 | EState VSA Descriptor 10 ( 9.17 <= x < 15.00) | MOE type descriptors using EState indices and surface area contribution |
| 13 | TopoPSA(NO) | Topological polar surface area (use only nitrogen and oxygen) | Topological polar surface area descriptor |
| 14 | mZagreb1 | modified Zagreb index (version 1) | Zagreb index descriptor |
| 15 | Kd_average | Average value of force constants of the dihedral angle | FF-related descriptor |
| 16 | Nd_average | Average value of multiplicity for the torsional angle parameters | FF-related descriptor |
| 17 | MW | Molecular weight of monomer | TC- related descriptor |
| 18 | MW_ratio | The ratio of the molecular weight of the main chain to that of the monomer | TC- related descriptor |
| 19 | Vdw | Van der Waals volume of the monomer | TC- related descriptor |
| 20 | Cross-sectional | Equivalent cross-sectional area of polymer chain | TC- related descriptor |

**Section S4. Construction of different ML models with optimized descriptors**

Figure S2 shows the RF, XGBoost and MLP models trained with the optimized descriptors. The test set is consisted of 100 randomly selected polymer structures from the benchmark dataset.

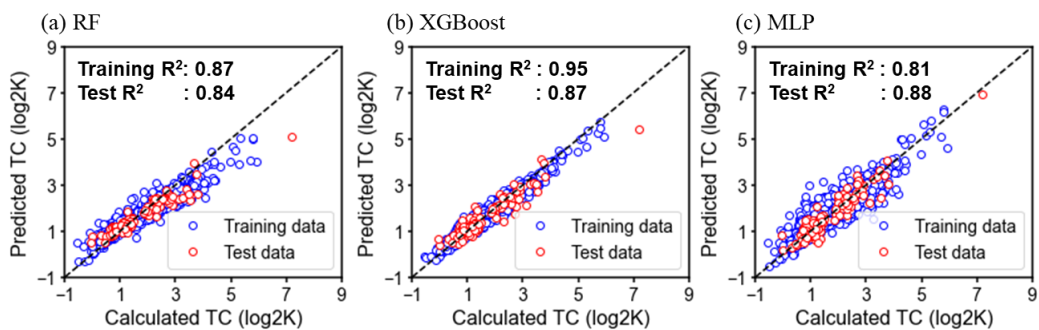

**Fig. S2.** Construction of different ML models with optimized descriptors

**Section S5. Accuracy of ML models at different descriptors downselection stages**

Figure S3 shows the accuracy of ML models at different descriptors downselection stages. The extra principal component analysis (PCA) [2] technique with 95% variance was used to compare.

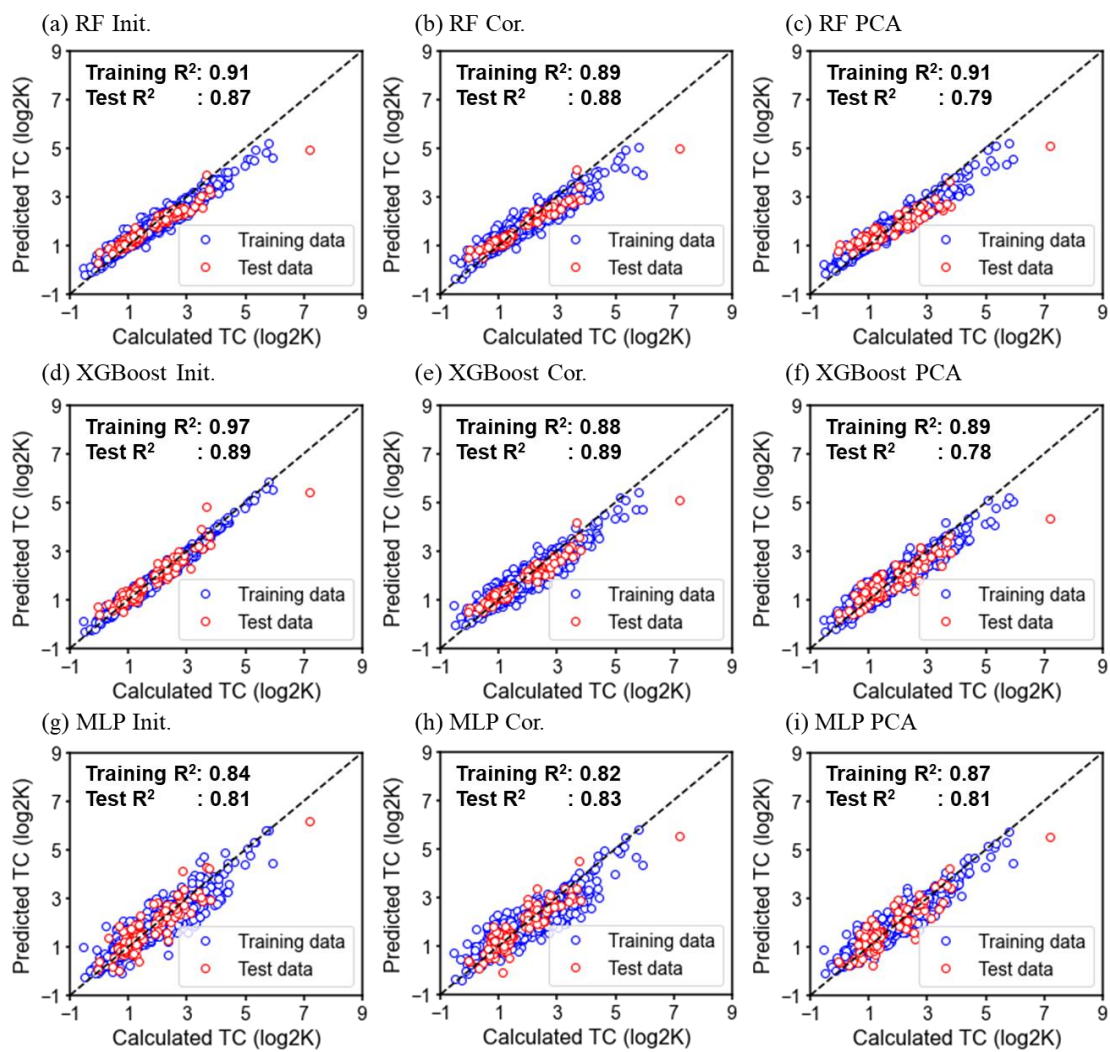

**Fig. S3.** Accuracy of ML models at different descriptors downselection stages

Figure S4 represents the relationship between the number of principal components and cumulative variance. Here, the PCA with 19 principal components of 95% variance was performed.

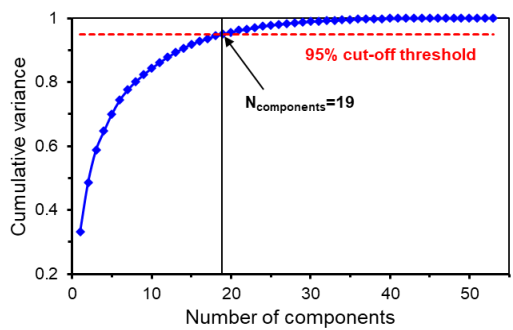

**Fig. S4.** Relationship between the number of principal components and cumulative variance

**Section S6. Accuracy of ML models with various graph descriptors**

Figure S5 shows the training and testing accuracy of ML models using Mol2vec [3], MACCS [4], Morgan and Morgan count (cMorgan) [5] graph descriptors.

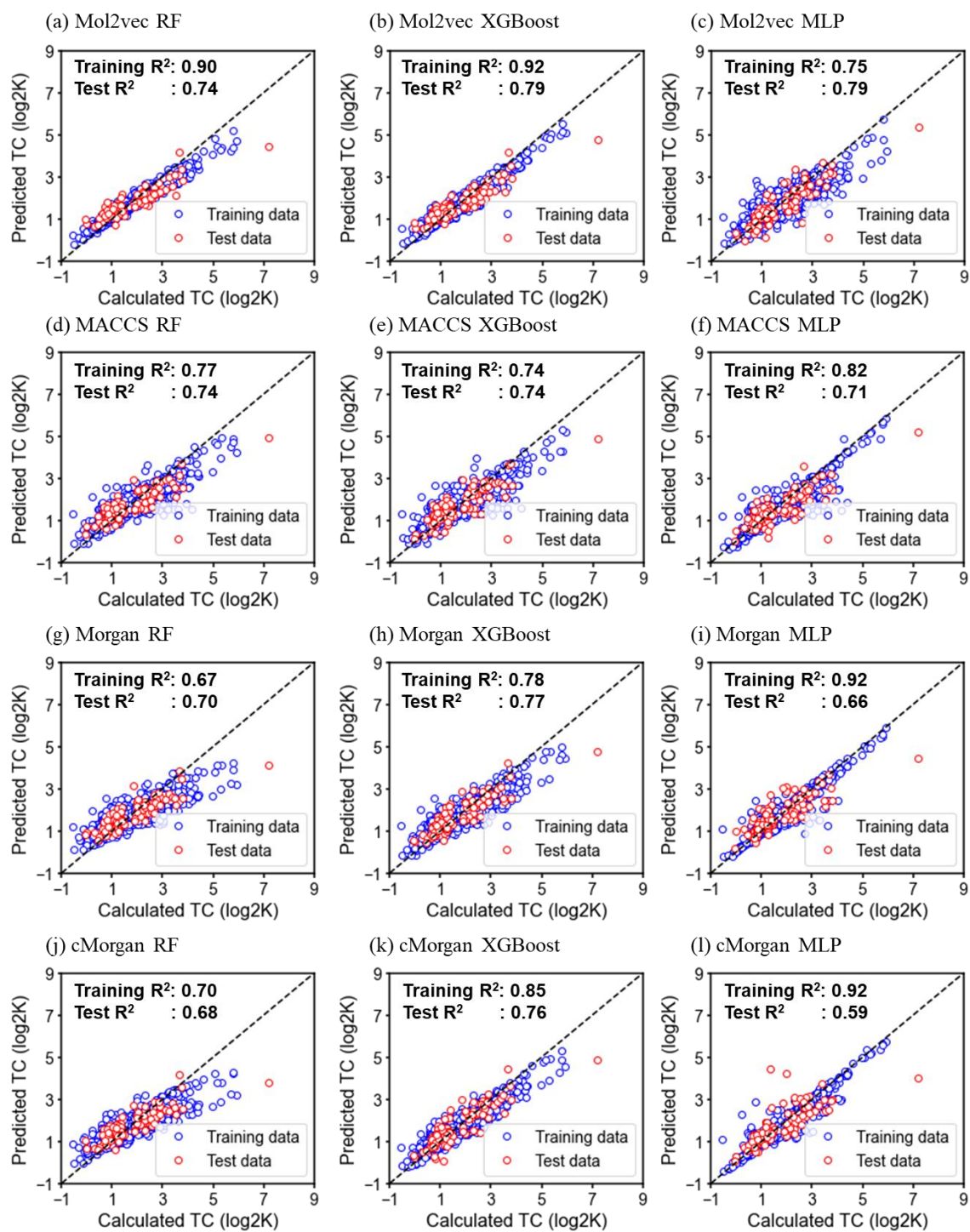

**Fig. S5.** Accuracy of ML models with various graph descriptors

**Section S7. Analysis of the feature importance**

Figure S6 exhibits the SHAP values [6] for each optimized descriptor of the training data set polymers as a function of corresponding descriptor value. The subplots are sorted by the average SHAP value. The SHAP values of the top-ranked features corresponding to the training dataset polymers show a monotonic relationship with the feature values overall, while the lower-ranked features have difficulty in capturing this rule. Wherein, the relationships between two important MD-inspired descriptors and TC are shown in Fig. S7.

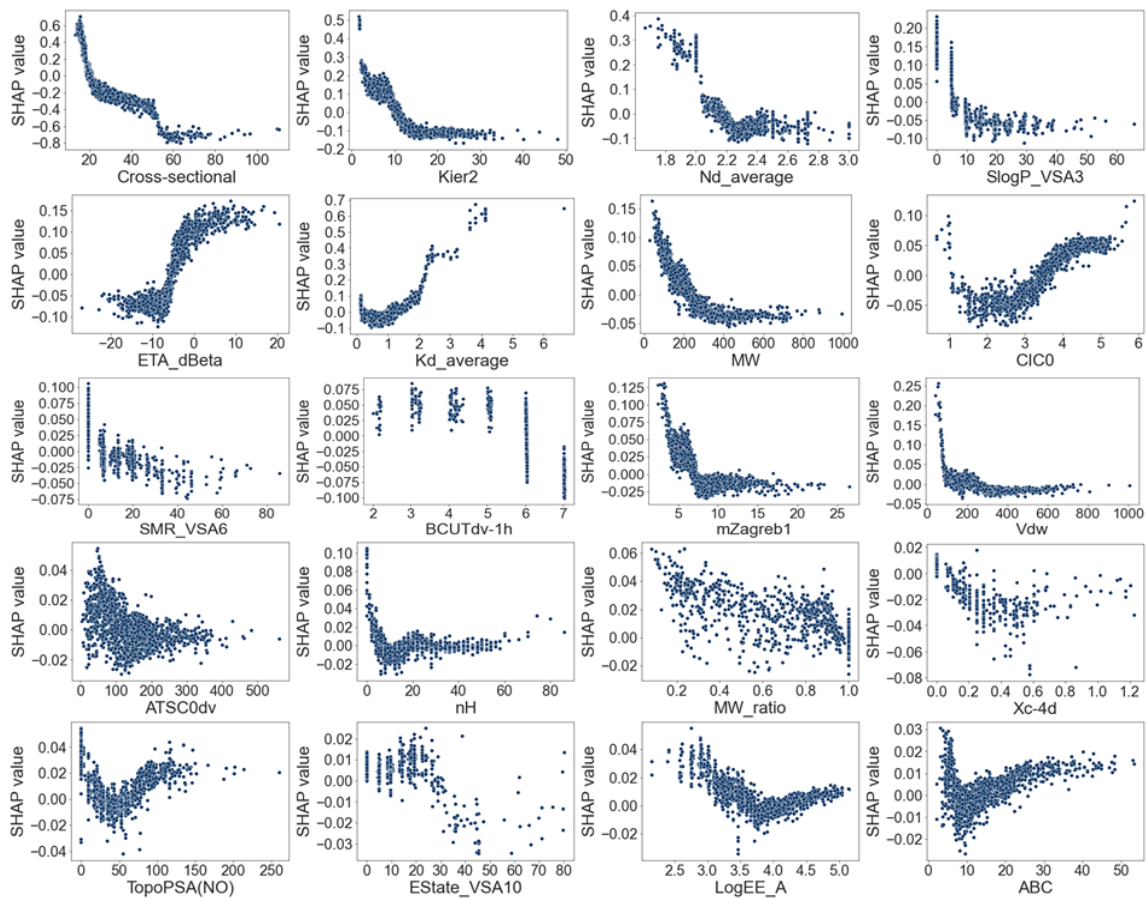

**Fig. S6.** Distribution of SHAP values for each optimized descriptor

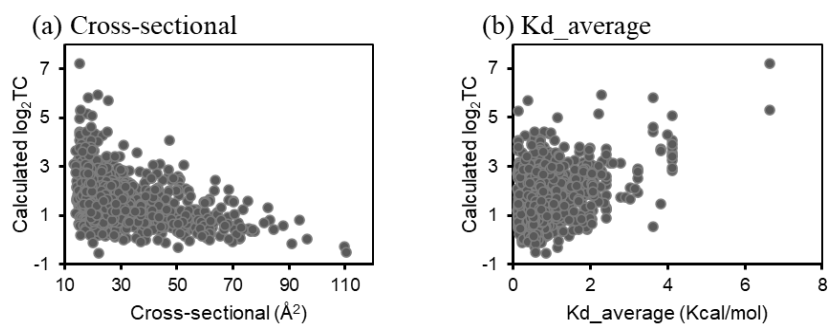

**Fig. S7.** Relationships between two important MD-inspired descriptors and TC of 1735 benchmark polymers.

## Section S8. Discovery of polymers with high thermal conductivity (PHTC) by proposed ML workflow

We performed virtual screening of polymer data in PolyInfo and PI1M based on the trained RF, XGBoost and MLP machine learning models, respectively. In this work, totally of 107 polymer structures with TC greater than 20 W/mK were identified and validated by MD simulations, which are listed in Table S3. In this work, a total of 107 polymer structures with thermal conductivity greater than 20 W/mK were identified and validated by molecular dynamics simulations, as listed in Table 4, and the relevant monomer structure can be viewed in Figure S8 by polymer ID (PID). The TC obtained from RF, XGBoost and MLP model predictions and MD simulations are $\log_2$TC in W/mK. The effective cross-sectional area of the polymer is given in $\text{Å}^2$.

**Table S4.** MD-validated polymer structures with thermal conductivity greater than 20 W/mK

| PID | SMILES | SA score | Cross-sectional | RF | XGBoost | MLP | MD |
|---|---|---|---|---|---|---|---|
| PHTC001 | [*]c1ccc([*])cc1 | 1.05 | 18.32 | 5.00 | 5.77 | 6.28 | 5.81 |
| PHTC002 | [*]CC[*] | 1.12 | 15.69 | 3.91 | 4.66 | 5.30 | 5.28 |
| PHTC003 | [*]c1cccc([*])c1 | 1.50 | 20.74 | 4.02 | 5.21 | 5.86 | 4.59 |
| PHTC004 | [*]CC(=O)N[*] | 2.07 | 15.13 | 4.00 | 4.34 | 4.05 | 4.41 |
| PHTC005 | [*]c1cccc([*])n1 | 2.15 | 19.72 | 4.65 | 4.98 | 5.65 | 5.06 |
| PHTC006 | [*]C=Cc1ccc([*])cc1 | 2.17 | 16.76 | 4.29 | 4.36 | 5.48 | 5.13 |
| PHTC007 | [*]c1ccc2cc([*])ccc2c1 | 2.17 | 18.28 | 3.55 | 4.37 | 5.01 | 5.12 |
| PHTC008 | [*]CC(=O)O[*] | 2.21 | 15.09 | 4.01 | 4.45 | 4.36 | 4.98 |
| PHTC009 | [*]c1cccc2c([*])cccc12 | 2.22 | 23.86 | 3.17 | 3.81 | 4.36 | 4.35 |
| PHTC010 | [*]c1ccc([*])c(Cl)c1 | 2.23 | 21.89 | 3.73 | 4.06 | 5.34 | 4.61 |
| PHTC011 | [*]c1ccc2ccc([*])cc2c1 | 2.28 | 19.21 | 3.81 | 4.49 | 5.05 | 4.62 |
| PHTC012 | [*]OCC([*])=O | 2.28 | 20.86 | 3.31 | 4.17 | 3.70 | 4.33 |
| PHTC013 | [*]c1ccc([*])nc1 | 2.33 | 16.98 | 4.85 | 5.08 | 5.45 | 4.50 |
| PHTC014 | [*]c1ccc([*])nn1 | 2.48 | 15.68 | 4.87 | 5.91 | 4.97 | 5.05 |
| PHTC015 | [*]C(F)(F)C([*])(F)F | 2.51 | 25.57 | 4.05 | 5.27 | 5.07 | 5.71 |
| PHTC016 | [*]NC(=O)N[*] | 2.57 | 14.34 | 4.53 | 4.91 | 3.76 | 5.47 |
| PHTC017 | [*]c1cnc([*])cn1 | 2.62 | 15.75 | 5.11 | 6.08 | 5.85 | 6.62 |
| PHTC018 | [*]c1ccc2nc([*])ccc2n1 | 2.64 | 16.79 | 3.96 | 4.00 | 5.62 | 5.93 |
| PHTC019 | [*]Oc1ccc([*])nc1 | 2.65 | 14.79 | 4.16 | 4.46 | 4.17 | 4.43 |
| PHTC020 | [*]Oc1ccc([*])cn1 | 2.65 | 14.74 | 4.16 | 4.52 | 4.17 | 4.76 |
| PHTC021 | [*]c1ccc2nc([*])ccc2c1 | 2.78 | 17.62 | 3.96 | 4.19 | 5.59 | 4.38 |
| PHTC022 | [*]c1cncc([*])n1 | 2.80 | 18.06 | 5.08 | 5.84 | 5.53 | 4.34 |
| PHTC023 | [*]c1ccnc([*])n1 | 2.82 | 18.06 | 5.08 | 5.82 | 5.52 | 5.00 |
| PHTC024 | [*]NNC([*])=O | 2.86 | 14.28 | 4.19 | 4.56 | 3.32 | 4.83 |

| PID | SMILES | SA score | Cross-sectional | RF | XGBoost | MLP | MD |
|---|---|---|---|---|---|---|---|
| PHTC025 | [*]c1csc([*])n1 | 2.87 | 15.82 | 5.11 | 6.20 | 5.76 | 5.75 |
| PHTC026 | [*]c1cnc2cc([*])ccc2c1 | 2.91 | 17.29 | 3.99 | 4.24 | 5.63 | 4.82 |
| PHTC027 | [*]c1ccc2oc([*])nc2c1 | 2.97 | 15.17 | 3.91 | 3.55 | 3.59 | 4.34 |
| PHTC028 | [*]c1ccc(-c2ccc([*])s2)s1 | 3.00 | 15.98 | 4.06 | 4.26 | 4.22 | 5.64 |
| PHTC029 | [*]C=Cc1ccc([*])c(F)c1 | 3.04 | 17.60 | 3.89 | 4.34 | 4.80 | 4.41 |
| PHTC030 | [*]C=C/C=C\[*] | 3.07 | 15.68 | 5.06 | 5.35 | 5.79 | 5.33 |
| PHTC031 | [*]C=C/C=C/[*] | 3.07 | 14.06 | 5.05 | 5.10 | 5.97 | 7.55 |
| PHTC032 | [*]/C=C/C=C/[*] | 3.07 | 14.26 | 5.05 | 5.10 | 5.95 | 6.91 |
| PHTC033 | [*]c1nc([*])nc(Cl)n1 | 3.09 | 13.85 | 4.53 | 5.64 | 5.63 | 6.51 |
| PHTC034 | [*]C=C[*] | 3.12 | 15.35 | 5.08 | 5.44 | 6.92 | 7.21 |
| PHTC035 | [*]C=C/C=C/C=C/[*] | 3.20 | 13.73 | 4.65 | 4.34 | 4.86 | 7.42 |
| PHTC036 | [*]/C=C/C=C/C=C/[*] | 3.20 | 13.60 | 4.65 | 4.34 | 4.87 | 7.31 |
| PHTC037 | [*]c1nnc([*])s1 | 3.27 | 8.76 | 5.09 | 6.04 | 6.14 | 5.75 |
| PHTC038 | [*]CCOC(=O)NCCNC(=O)O[*] | 3.33 | 15.55 | 3.20 | 4.10 | 3.65 | 4.38 |
| PHTC039 | [*]C(=O)c1cnc([*])cn1 | 3.34 | 15.80 | 4.11 | 4.34 | 4.35 | 4.54 |
| PHTC040 | [*]c1cnc([*])s1 | 3.35 | 15.34 | 4.86 | 5.77 | 5.40 | 4.95 |
| PHTC041 | [*]C1=Cc2ccc([*])cc21 | 3.37 | 18.78 | 4.37 | 4.64 | 5.50 | 5.11 |
| PHTC042 | [*]OC(=O)O[*] | 3.38 | 11.93 | 3.97 | 4.05 | 3.96 | 5.61 |
| PHTC043 | [*]c1ccc2nc([*])ncc2n1 | 3.39 | 15.82 | 3.98 | 3.95 | 5.54 | 5.83 |
| PHTC044 | [*]c1cc(Cl)c([*])nn1 | 3.45 | 19.29 | 4.19 | 5.08 | 4.97 | 4.52 |
| PHTC045 | [*]c1cnc([*])nn1 | 3.49 | 14.33 | 5.06 | 5.97 | 5.57 | 6.46 |
| PHTC046 | [*]c1nnc([*])c(Cl)c1Cl | 3.51 | 15.39 | 4.11 | 5.12 | 5.41 | 4.60 |
| PHTC047 | [*]c1cnnc([*])n1 | 3.53 | 16.38 | 5.04 | 5.92 | 5.33 | 5.60 |
| PHTC048 | [*]c1cnc([*])n1C | 3.55 | 19.00 | 4.25 | 4.46 | 5.63 | 4.49 |
| PHTC049 | [*]c1ccc(-c2nnc([*])s2)s1 | 3.63 | 14.53 | 4.15 | 4.39 | 4.83 | 5.39 |
| PHTC050 | [*]c1nnc([*])c(F)c1F | 3.64 | 12.41 | 4.40 | 5.32 | 4.51 | 4.75 |
| PHTC051 | [*]c1ccc(-c2cnc([*])s2)s1 | 3.76 | 15.25 | 4.17 | 4.44 | 4.85 | 5.51 |
| PHTC052 | [*]C(Cl)=C([*])Cl | 3.86 | 20.41 | 4.17 | 4.38 | 4.81 | 5.13 |
| PHTC053 | [*]/C(Cl)=C(/[*])Cl | 3.86 | 20.56 | 4.13 | 4.42 | 4.80 | 5.08 |
| PHTC054 | [*]C#C[*] | 3.87 | 5.85 | 4.23 | 4.58 | 5.15 | 9.63 |
| PHTC055 | [*]C1=CC=C1[*] | 3.90 | 22.30 | 4.36 | 5.72 | 5.65 | 4.62 |
| PHTC056 | [*]c1nnc([*])c(O)n1 | 3.90 | 16.45 | 4.50 | 5.18 | 4.20 | 4.96 |
| PHTC057 | [*]c1noc([*])c1Br | 3.90 | 11.67 | 4.61 | 5.55 | 4.77 | 4.97 |
| PHTC058 | [*]C(=O)N1C(=O)N([*])C1=O | 4.04 | 16.85 | 3.66 | 4.36 | 5.49 | 5.37 |
| PHTC059 | [*]NC(=O)C(=S)C([*])=O | 4.08 | 21.85 | 4.01 | 5.07 | 4.60 | 5.91 |
| PHTC060 | [*]C(=O)c1nnc([*])nn1 | 4.09 | 11.89 | 4.36 | 5.35 | 4.15 | 5.95 |
| PHTC061 | [*]c1cc2nc([*])sc2nn1 | 4.09 | 14.51 | 4.38 | 4.81 | 5.11 | 4.35 |
| PHTC062 | [*]C(=O)C([*])=S | 4.20 | 16.81 | 3.78 | 4.92 | 4.41 | 5.03 |
| PHTC063 | [*]C1=C(F)C(Cl)=C1[*] | 4.28 | 17.13 | 4.69 | 4.59 | 5.00 | 5.84 |

| PID | SMILES | SA score | Cross-sectional | RF | XGBoost | MLP | MD |
|---|---|---|---|---|---|---|---|
| PHTC064 | [*]C1=CC=C2C([*])=CC=C12 | 4.32 | 22.02 | 3.94 | 3.73 | 4.63 | 6.16 |
| PHTC065 | [*]NC(=O)C([*])=N | 4.32 | 18.67 | 3.86 | 5.17 | 3.12 | 4.37 |
| PHTC066 | [*]SC([*])=O | 4.34 | 8.22 | 4.70 | 5.29 | 4.52 | 6.73 |
| PHTC067 | [*]/C=C/C=C([*])Cl | 4.36 | 17.41 | 4.65 | 3.72 | 4.63 | 5.90 |
| PHTC068 | [*]/C(F)=C(/[*])C#N | 4.36 | 20.54 | 3.77 | 5.11 | 3.47 | 6.02 |
| PHTC069 | [*]C=Cc1csc(-c2cc([*])cs2)c1 | 4.45 | 15.88 | 3.69 | 3.61 | 3.08 | 4.33 |
| PHTC070 | [*]/C=C/C=C/C=C([*])Cl | 4.45 | 15.96 | 4.28 | 4.78 | 3.99 | 5.66 |
| PHTC071 | [*]/C=C/C=C/C=C(/[*])Cl | 4.45 | 15.96 | 4.28 | 4.78 | 3.99 | 5.66 |
| PHTC072 | [*]/C(F)=C(/[*])Cl | 4.46 | 18.39 | 4.44 | 5.10 | 4.90 | 6.18 |
| PHTC073 | [*]C(F)=C([*])Cl | 4.46 | 17.50 | 4.69 | 5.25 | 4.94 | 5.83 |
| PHTC074 | [*]OC([*])=S | 4.53 | 14.49 | 4.63 | 5.07 | 3.76 | 5.20 |
| PHTC075 | [*]C#CC=C[*] | 4.54 | 18.03 | 4.51 | 5.02 | 5.18 | 5.16 |
| PHTC076 | [*]OC([*])=C | 4.54 | 19.19 | 4.07 | 4.36 | 3.57 | 4.55 |
| PHTC077 | [*]/C=C(/[*])Cl | 4.55 | 21.62 | 4.28 | 4.80 | 5.14 | 6.03 |
| PHTC078 | [*]C=C([*])C#N | 4.58 | 24.93 | 3.53 | 4.00 | 2.70 | 4.88 |
| PHTC079 | [*]C(=S)C(=O)C(F)(F)C([*])(F)F | 4.59 | 25.30 | 3.19 | 4.28 | 3.92 | 4.41 |
| PHTC080 | [*]SS[*] | 4.66 | 13.04 | 3.77 | 3.91 | 0.95 | 4.93 |
| PHTC081 | [*]/C=C/C=C/C=C([*])N | 4.66 | 15.44 | 4.20 | 4.41 | 4.69 | 5.53 |
| PHTC082 | [*]NNC([*])=N | 4.67 | 15.08 | 4.45 | 4.61 | 4.41 | 4.51 |
| PHTC083 | [*]C1=CC2=NC([*])=CC2=N1 | 4.75 | 17.77 | 4.55 | 4.07 | 4.51 | 6.10 |
| PHTC084 | [*]C#CC#CC=C[*] | 4.78 | 12.75 | 4.11 | 3.79 | 4.61 | 5.99 |
| PHTC085 | [*]C=C([*])F | 4.82 | 17.93 | 4.71 | 5.35 | 4.81 | 6.64 |
| PHTC086 | [*]/C=C/N[*] | 4.83 | 15.68 | 5.02 | 5.28 | 6.52 | 5.44 |
| PHTC087 | [*]C#C/C(C#N)=C(/[*])C#N | 4.88 | 15.14 | 3.46 | 5.00 | 4.71 | 5.37 |
| PHTC088 | [*]/N=C/C(=S)N[*] | 4.88 | 18.77 | 4.37 | 4.29 | 5.00 | 5.06 |
| PHTC089 | [*]/N=N/C([*])=O | 4.90 | 8.82 | 4.55 | 5.76 | 6.04 | 7.35 |
| PHTC090 | [*]/C=C/O[*] | 4.92 | 15.71 | 3.94 | 3.83 | 3.43 | 4.60 |
| PHTC091 | [*]=CC=NN=[*] | 4.98 | 5.95 | 4.69 | 4.92 | 7.53 | 4.71 |
| PHTC092 | [*]NN[*] | 5.00 | 11.88 | 4.79 | 5.37 | 6.33 | 4.60 |
| PHTC093 | [*]/N=N/[*] | 5.05 | 5.21 | 4.83 | 5.95 | 6.81 | 10.38 |
| PHTC094 | [*]N=N[*] | 5.05 | 5.93 | 4.83 | 5.95 | 6.76 | 10.01 |
| PHTC095 | [*]/N=C(/[*])Cl | 5.08 | 9.62 | 5.06 | 5.71 | 6.50 | 6.98 |
| PHTC096 | [*]C#C/C(C#N)=C(/[*])Cl | 5.14 | 14.06 | 3.63 | 5.06 | 4.58 | 5.79 |
| PHTC097 | [*]/C=N/C=C(/[*])C | 5.14 | 17.30 | 4.02 | 3.64 | 4.36 | 4.72 |
| PHTC098 | [*]/C=C/C=CS[*] | 5.17 | 14.14 | 4.14 | 3.28 | 4.13 | 4.36 |
| PHTC099 | [*]C1=CC2=CC([*])=NC2=N1 | 5.28 | 17.50 | 4.62 | 4.10 | 4.55 | 5.79 |
| PHTC100 | [*]/C=C/N=C([*])Cl | 5.38 | 19.65 | 4.16 | 4.37 | 4.61 | 5.60 |
| PHTC101 | [*]/C=C/S[*] | 5.44 | 14.40 | 4.57 | 4.20 | 5.74 | 4.35 |
| PHTC102 | [*]/C=N/[*] | 5.53 | 14.02 | 5.10 | 6.07 | 6.07 | 6.82 |

| PID | SMILES | SA score | Cross-sectional | RF | XGBoost | MLP | MD |
|---|---|---|---|---|---|---|---|
| PHTC103 | [*]/N=N/N=C(/[*])C#N | 5.66 | 11.24 | 3.69 | 5.07 | 4.33 | 6.68 |
| PHTC104 | [*]/C=N/C=C/C=C(/[*])Cl | 5.70 | 15.50 | 4.33 | 4.46 | 4.07 | 5.70 |
| PHTC105 | [*]/C=C(/[*])[O-] | 5.73 | 18.54 | 4.07 | 4.66 | 2.87 | 6.10 |
| PHTC106 | [*]/C=C/C=C/N=N/[*] | 5.76 | 12.71 | 4.74 | 4.11 | 7.01 | 6.59 |
| PHTC107 | [*]C([2H])=C([*])[2H] | 7.27 | 11.30 | 4.32 | 4.33 | 1.07 | 7.48 |

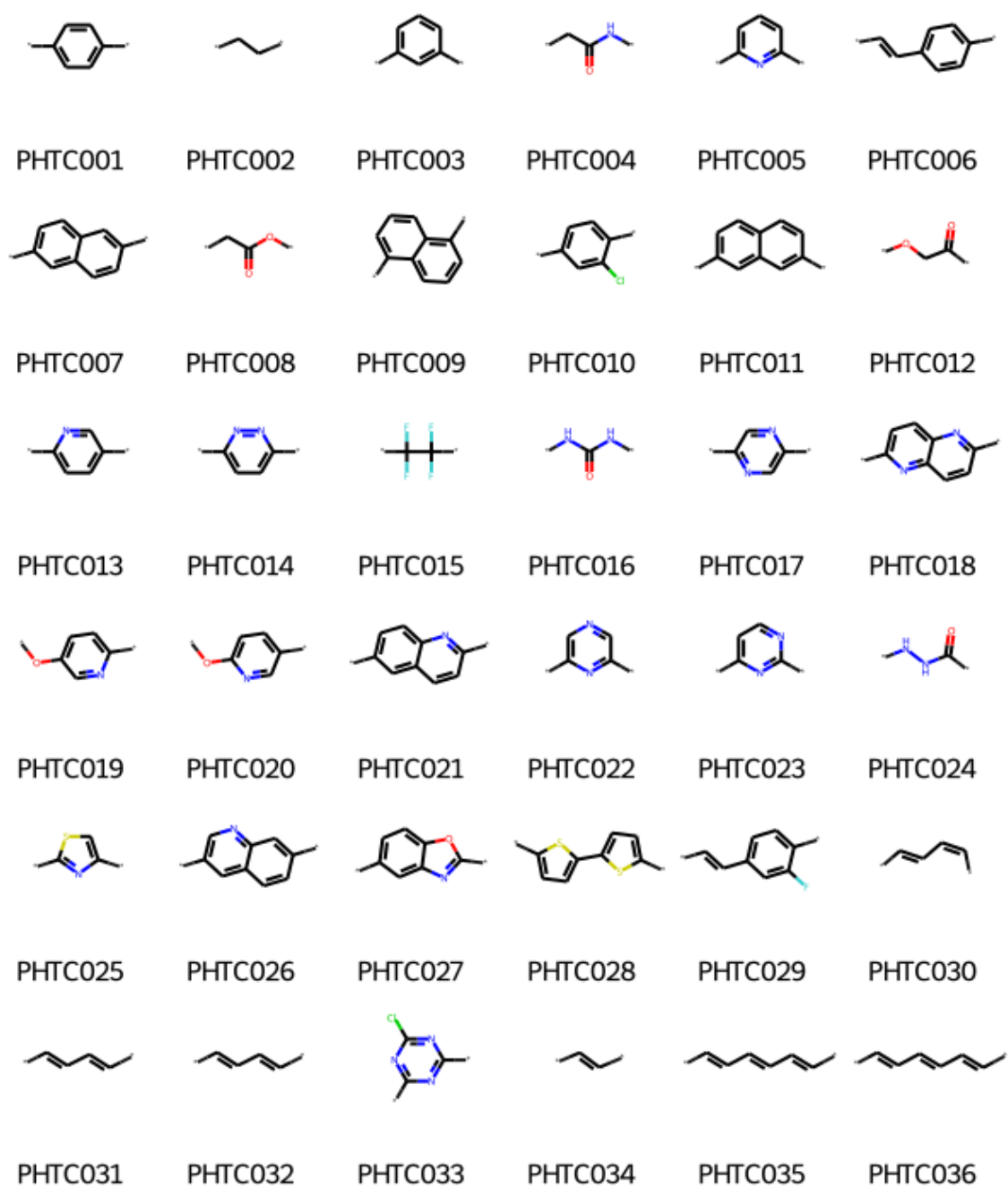

PHTC001 PHTC002 PHTC003 PHTC004 PHTC005 PHTC006

PHTC007 PHTC008 PHTC009 PHTC010 PHTC011 PHTC012

PHTC013 PHTC014 PHTC015 PHTC016 PHTC017 PHTC018

PHTC019 PHTC020 PHTC021 PHTC022 PHTC023 PHTC024

PHTC025 PHTC026 PHTC027 PHTC028 PHTC029 PHTC030

PHTC031 PHTC032 PHTC033 PHTC034 PHTC035 PHTC036

**Fig. S8.** Continued

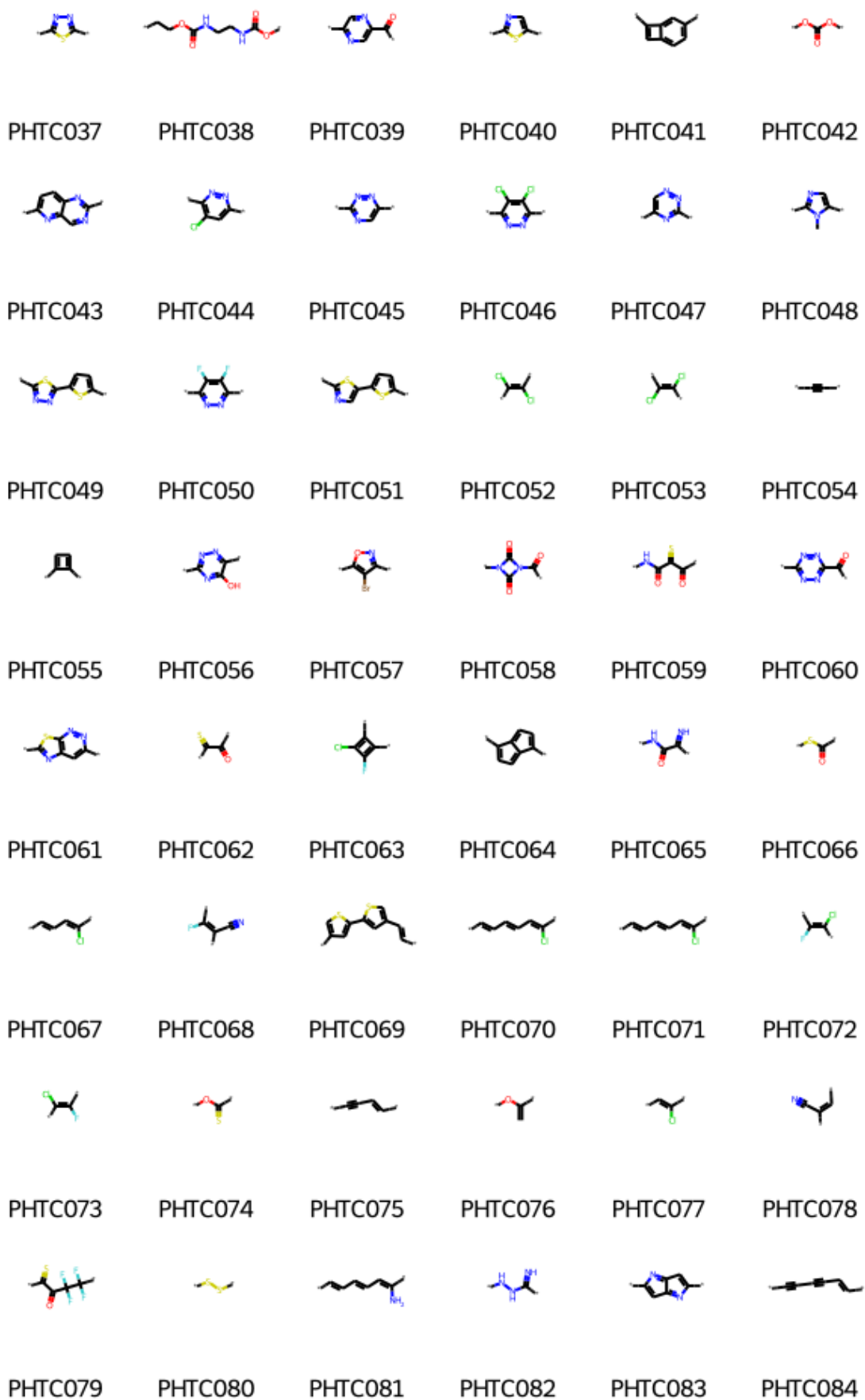

| | | | | | |
|---|---|---|---|---|---|
| PHTC037 | PHTC038 | PHTC039 | PHTC040 | PHTC041 | PHTC042 |
| PHTC043 | PHTC044 | PHTC045 | PHTC046 | PHTC047 | PHTC048 |
| PHTC049 | PHTC050 | PHTC051 | PHTC052 | PHTC053 | PHTC054 |
| PHTC055 | PHTC056 | PHTC057 | PHTC058 | PHTC059 | PHTC060 |
| PHTC061 | PHTC062 | PHTC063 | PHTC064 | PHTC065 | PHTC066 |
| PHTC067 | PHTC068 | PHTC069 | PHTC070 | PHTC071 | PHTC072 |
| PHTC073 | PHTC074 | PHTC075 | PHTC076 | PHTC077 | PHTC078 |
| PHTC079 | PHTC080 | PHTC081 | PHTC082 | PHTC083 | PHTC084 |

**Fig. S8.** Continued

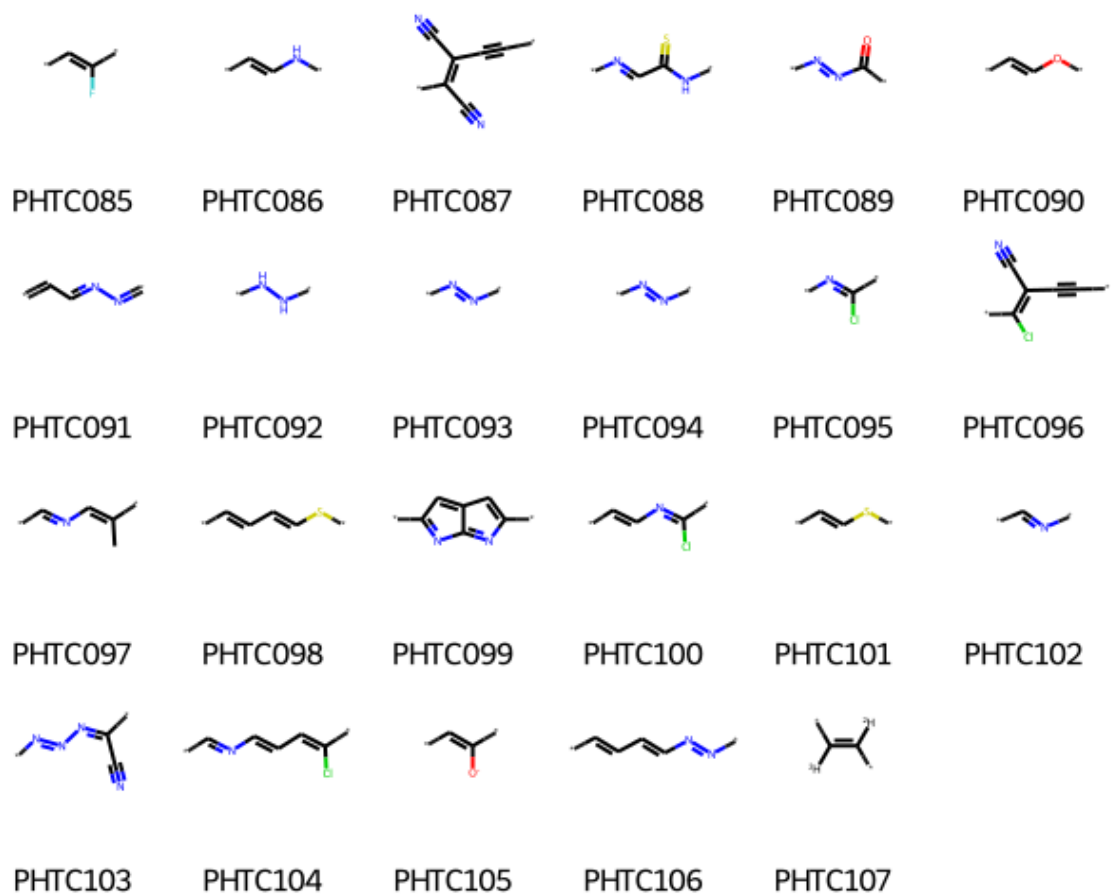

| | | | | | |
|---|---|---|---|---|---|
| PHTC085 | PHTC086 | PHTC087 | PHTC088 | PHTC089 | PHTC090 |
| PHTC091 | PHTC092 | PHTC093 | PHTC094 | PHTC095 | PHTC096 |
| PHTC097 | PHTC098 | PHTC099 | PHTC100 | PHTC101 | PHTC102 |
| PHTC103 | PHTC104 | PHTC105 | PHTC106 | PHTC107 | |

**Fig. S8.** Polymer monomer structures with TC greater than 20 W/mK verified by MD simulations. Polymer properties can be found in Table S4 using the PID.

**Section S9. Symbolic regression for characterizing the TC of promising polymers**

The 107 promising polymer structures (TC > 20 W/mK) with optimized descriptors were utilized for symbolic regression (SR), where 20% structures were randomly selected for testing set. The mathematical formulae were acquired and selected using an efficient stepwise strategy with SR based on a genetic programming (GPSR) as implemented in the gplearn code [7]. At first, Pearson coefficients were used as evaluation metrics of training fitness to filter optimized descriptors and generate sub-descriptors, and the grid search strategy with the hyperparameters listed in Table S5 was applied in GPSR. The hyperparametric combinations corresponding to the 22380 formulas are characterized by complexity and training fitness of Pearson coefficients (PC) as shown in Fig. S9a. The complexity is equivalent to the formula length in gplearn. Among these formulas, 4365 of them with PC values not less than 0.85 are statistically by complexity in Fig. S9b. Generally, the formulas with large complexity have a higher training fitness. But the formulas are also usually quite long and do not facilitate the calculation and analysis. Only formulas with large fitness and low complexity are appropriate, so we selected 158 formulas with the PC value not less than 0.85 and complexity within 10 for subsequent analysis. By counting the frequency of occurrence of the 20 optimized descriptors in 158 formulas, the first 8 descriptors were finally retained, as presented in Fig. S9c. It is worth emphasizing that the MD-inspired descriptors of cross-sectional area (cross-sectional) and dihedral force constants (Kd_average) appeared in each of the formulas. And, we additionally estimated the frequency of occurrence of sub-descriptors created by taking the inverse $(-x)$, logarithm $(\ln x)$, and so on, as shown in Fig. S9d. As a result, a new ensemble containing 22 descriptors which have a strong association with TC is listed in Table S6 for retraining in gplearn.

Further, we reset the grid search hyperparameters listed in Table S7, and utilized the accuracy $R^2$ as the evaluation metric. The 19580 formulas with fitting accuracy greater than 0.6 versus complexity are visualized in Fig. S10a. The effective accuracy is the average of the training and testing accuracy. Despite the fact that the effective accuracy of some formulas exceeds 0.8, and their complexity is greater than 50. The Pareto front of accuracy $R^2$ vs. complexity (no more than 30) of 9073 mathematical formulas shown via density plot in Fig. S10b, and the points of c, d, e and f were identified by Latin hypercube sampling approach [8,9]. The formulas for the four points are listed in Table S8, and their fitting results compared with MD labeled $\log_2$TC are displayed in Fig. 10c-f. The definitions of the variables $x_0$ to $x_{21}$ can be found in Table S6.

**Table S5.** Setup of hyperparameters in gplearn for filter and generate new sub-descriptors

| Parameter | Value | Combination |
|---|---|---|
| Generations | 300 | 1 |
| Population size in every generation | 1000,2000 | 2 |
| Probability of crossover (pc) | [0.30,0.90], step=0.05 | |
| Subtree mutation (ps) | [(1-pc)/3,(1-pc)/2] (step= 0.01) | 746 |
| Hoist mutation (ph) | [(1-pc)/3,(1-pc)/2] (step = 0.01) | |
| Point mutation (pp) | 1-pc-ps-ph | |
| Function set | $\{+,-,\times,\div,\sqrt{x},\ln x,|x|,-x,1/x\}$ | 1 |
| Parsimony coefficient | auto | 1 |
| Metric | Pearson coefficient | 1 |
| Stopping criterial | 0.900 | 1 |
| Random_state | 0, 1, 2, 3, 4 | 5 |
| Init_depth | [2, 6], [4, 8], [6, 10] | 3 |

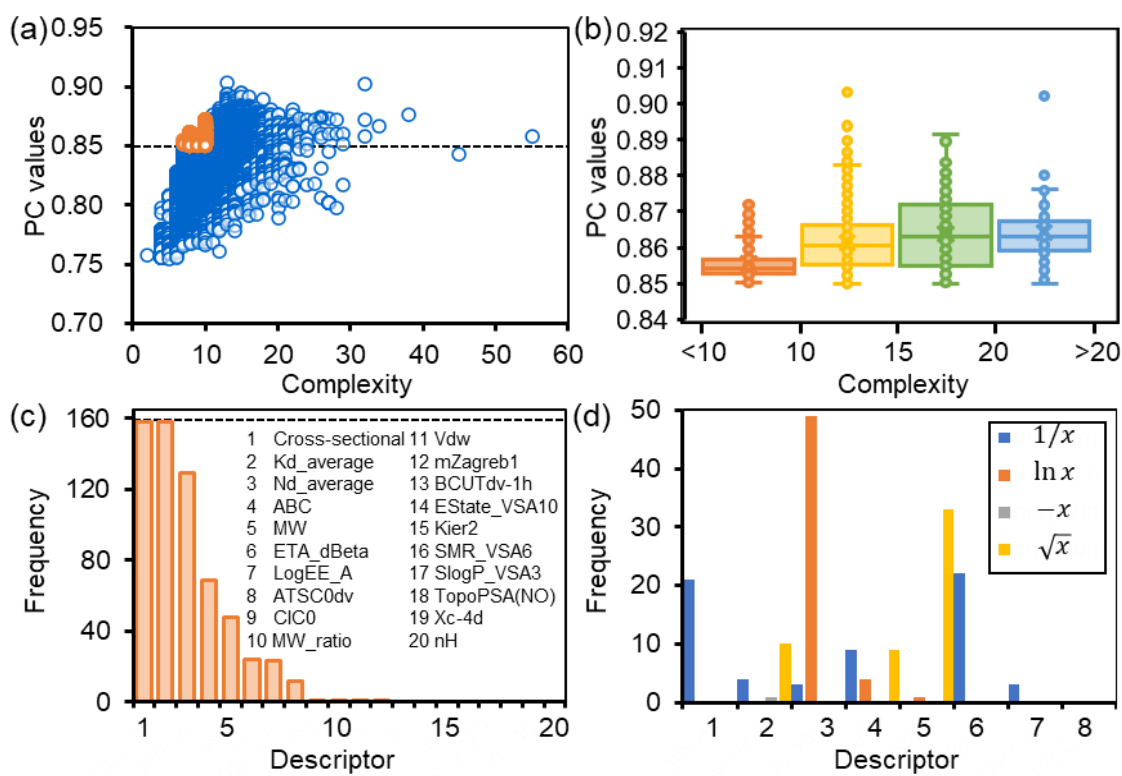

**Fig. S9.** Filtering optimized descriptors and creating sub-descriptors in gplearn through Pearson coefficient (PC) values. (a) Mathematical formula complexity versus PC values. (b) Statistics of formulas with Pearson coefficient not less than 0.85 by complexity. (c) and (d) Frequency of occurrence of optimized descriptors and sub-descriptors in 158 mathematical formulas (PC values >=0.85 and complexity <=10).

**Table S6.** New descriptors ensemble for GPSR

| Variable | Format | PC values | Variable | Format | PC values |
|---|---|---|---|---|---|
| $x_0$ | ABC | -0.37 | $x_{11}$ | -ETA_dBeta | 0.11 |
| $x_1$ | 10/ABC | 0.44 | $x_{12}$ | Kd_average | 0.20 |
| $x_2$ | $\sqrt{\text{ABC}}$ | -0.39 | $x_{13}$ | $\sqrt{\text{Kd\_average}}$ | 0.12 |
| $x_3$ | ATSC0dv | -0.32 | $x_{14}$ | LogEE_A | -0.40 |
| $x_4$ | Cross-sectional | -0.52 | $x_{15}$ | MW | -0.39 |
| $x_5$ | 10/Cross-sectional | 0.63 | $x_{16}$ | MW/100 | -0.39 |
| $x_6$ | 1-10/Cross-sectional | 0.59 | $x_{17}$ | 100/MW | 0.52 |
| $x_7$ | 100/Cross-sectional | 0.63 | $x_{18}$ | $\sqrt{\text{MW}}$ | -0.43 |
| $x_8$ | $\sqrt{100/\text{Cross} - \text{sectional}}$ | 0.62 | $x_{19}$ | Nd_average | -0.31 |
| $x_9$ | -10/Cross-sectional | -0.63 | $x_{20}$ | log Nd_average | -0.35 |
| $x_{10}$ | ETA_dBeta | -0.11 | $x_{21}$ | $-\log$ Nd_average | 0.44 |

**Table S7.** Reset of hyperparameters in gplearn for GPSR

| Parameter | Value | Combination |
|---|---|---|
| Generations | 300 | 1 |
| Population size in every generation | 5000 | 1 |
| Probability of crossover (pc) | [0.30,0.90], step=0.05 | |
| Subtree mutation (ps) | [(1-pc)/3,(1-pc)/2] (step= 0.01) | 746 |
| Hoist mutation (ph) | [(1-pc)/3,(1-pc)/2] (step = 0.01) | |
| Point mutation (pp) | 1-pc-ps-ph | |
| Function set | $\{+,-,\times,\div,\sqrt{x},\ln x\,,|x|,-x,1/x\}$ | 1 |
| Parsimony coefficient | 0.001, 0.003, 0.005 | 3 |
| Metric | $R^2$ | 1 |
| Stopping criterial | 0.900 | 1 |
| Random_state | 0, 1, 2, 3, 4 | 5 |
| Init_depth | [2, 6], [4, 8], [6, 10], [2, 10] | 4 |

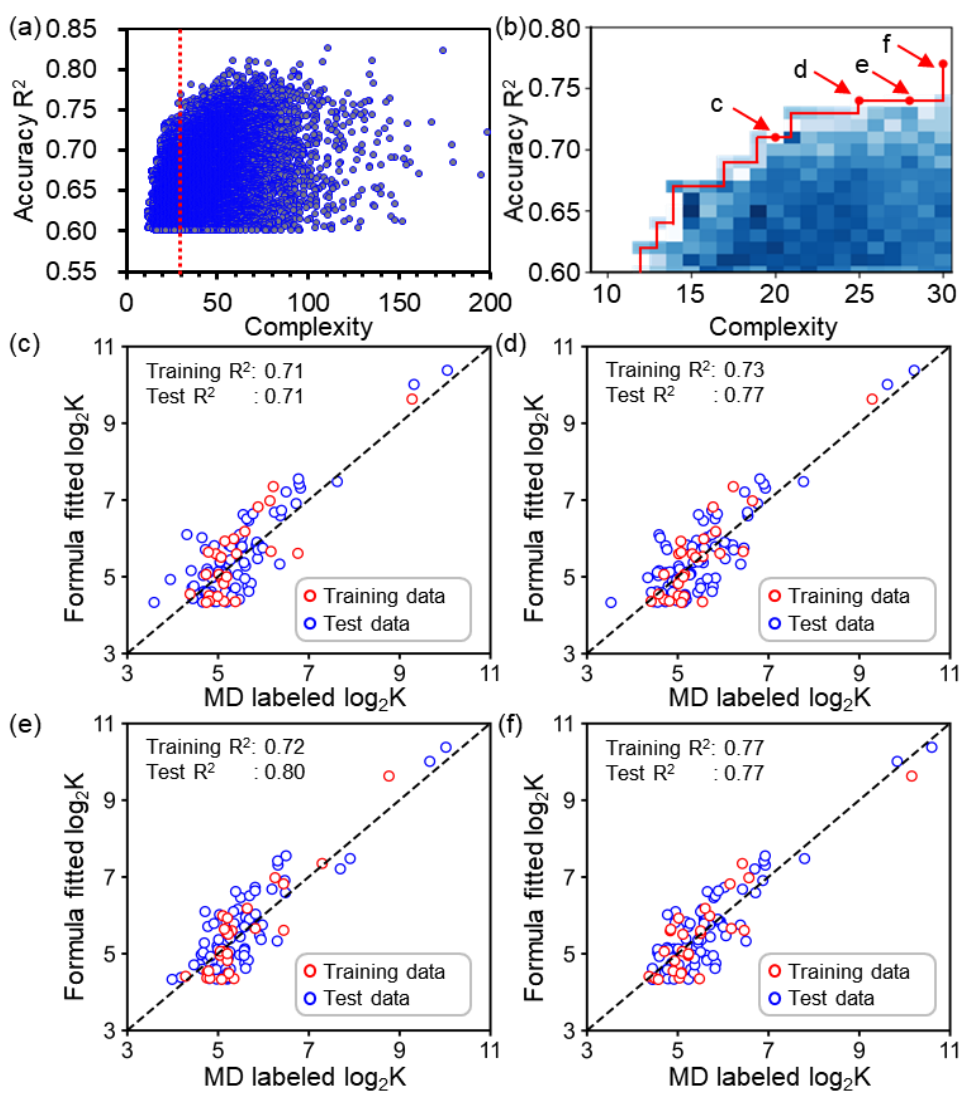

**Fig. S10.** GPSR for TC prediction of promising polymers. (a) 19580 formulas with fitting accuracy greater than 0.6 versus complexity. (b) Pareto front of accuracy R2 vs. complexity of 9073 mathematical formulas shown via density plot. The four points of c, d, e and f represent the four formulas at the Pareto front, whose fitting results vs. MD labeled TC are plotted in (c), (d), (e) and (f) respectively.

**Table S8.** The four mathematical formulas at the Pareto front in Fig. S10b

| Point | Formulas | $R^2$ | Complexity |
|---|---|---|---|
| $c$ | $log_2 K = \sqrt{(x_{12} - x_8 - x_{16} + 0.49) \times x_9 \times (x_6 + x_{20})} + \sqrt{ln(x_{12})}$ $+ x_{21}$ | 0.71 | 20 |
| $d$ | $log_2 K = \sqrt{x_{12} - ln\left[(x_8 - x_{11}) \times \dfrac{x_8}{ln(\sqrt{x_{19}})}\right] \times (x_6 + x_{20}) \times x_7}$ $+ x_{21} + \sqrt{ln\, x_{12}}$ | 0.74 | 25 |
| e | $log_2 K = ln\left[\dfrac{1}{x_{10} \times (x_1 - x_2)} - x_{12}\right] \times (x_5 - 0.824)$ $+ \dfrac{x_{17}}{\sqrt{\dfrac{x_3}{(x_{12} - x_{11}) \times x_5}}} + x_{21} + x_{13}$ | 0.74 | 28 |
| f | $log_2 K = \sqrt{x_8 + \left[\dfrac{x_{10}}{ln(ln\, x_{17} - x_{12}) + x_8} - x_{12}\right] \times (x_6 + x_{20}) \times \dfrac{x_8}{x_{20}}}$ $+ x_{21} + \sqrt{x_{19} - x_{16}}$ | 0.77 | 30 |